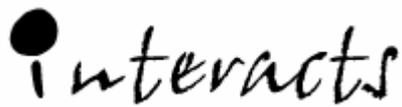

**Improving Interaction between NGOs, Universities, and Science Shops: Experiences and Expectations**

# COMMUNICATION OF SCIENCE SHOP MEDIATION:

# A Kaleidoscope of University-Society Relations

Loet Leydesdorff & Janelle Ward

February 2004

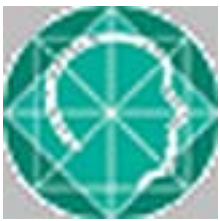
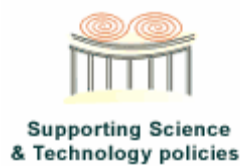

Supporting Science & Technology policies

A project funded by the European Commission/DG 12 under the Fifth RTD Framework Programme

Contract No. HPV1-CT-2001-60039

# Acknowledgement

We would like to thank our colleagues of the INTERACTS team for providing us with drafts of their case study reports and for their valuable comments on previous drafts of this report.

# Table of Contents

# 1.    Introduction

The Science Shop model was initiated in the Netherlands in the 1970s. During the 1980s, the model spread over Europe, but without much coordination. Part of the model is the modest scale of the operation. The crucial idea behind the Science Shops involves a working relationship between knowledge-producing institutions like universities and citizen groups that need relevant questions answered.  In providing this link, the relations between science and the public can be stimulated by providing such groups with access to the university and by offering active mediation of these questions. The costs of such an enterprise are relatively low given existing structures, and one can expect some synergetic surplus value on both sides when a match between public interest and research interests is reached.

Initial evaluations of the mediation and its effect on scientific output and curriculum formation were positive (Zaal & Leydesdorff, 1987) although it was also noted that the impacts can be asymmetrical for the parties involved (Leydesdorff & Van den Besselaar, 1988a and b). For example, a science shop project may provide researchers with access to a specific domain or even with a methodological challenge, while the clients may be interested mainly in the empirical results. Science shop questions can sometimes be used for the legitimation of new lines of research.

Given the local character of Science Shop activities, it can be expected after a few decades that the Science Shops have created their own "best practices."  However, these often vary considerably among one another given differences in relevant environments. For example, Science Shops can be more or less identified with interests of one of the parties involved: the clients, the social legitimation and embeddedness of university research, or the interests of students in higher education.

The scope of possible relations with clients can be expected to vary among disciplines, since disciplines entertain very different relations with non-scientific audiences. For example, in medicine a relation between the laboratory and clinical research is institutionally organized in the academic hospital, while other sciences communicate with relevant audiences through national research networks or through potentially global communication systems like the scientific journals. Internally, the sciences are differentiated in terms of their communication structures. Quality control of the communication is controlled by reputational structures (Whitley, 1984). These higher-level, but internal control mechanisms are likely to affect the scientific output of the research institution to an extent larger than its communication with the social environment. The internal communications are needed to legitimate the use of scientific results in applied research and in relations with clients. The underlying processes of knowledge production, however, may have been black boxed in the stage of external communication of the results and of application.

The mechanisms of quality control and diffusion are not to be considered as one-way streams, but as communication networks. This is schematically illustrated in Figure 1.

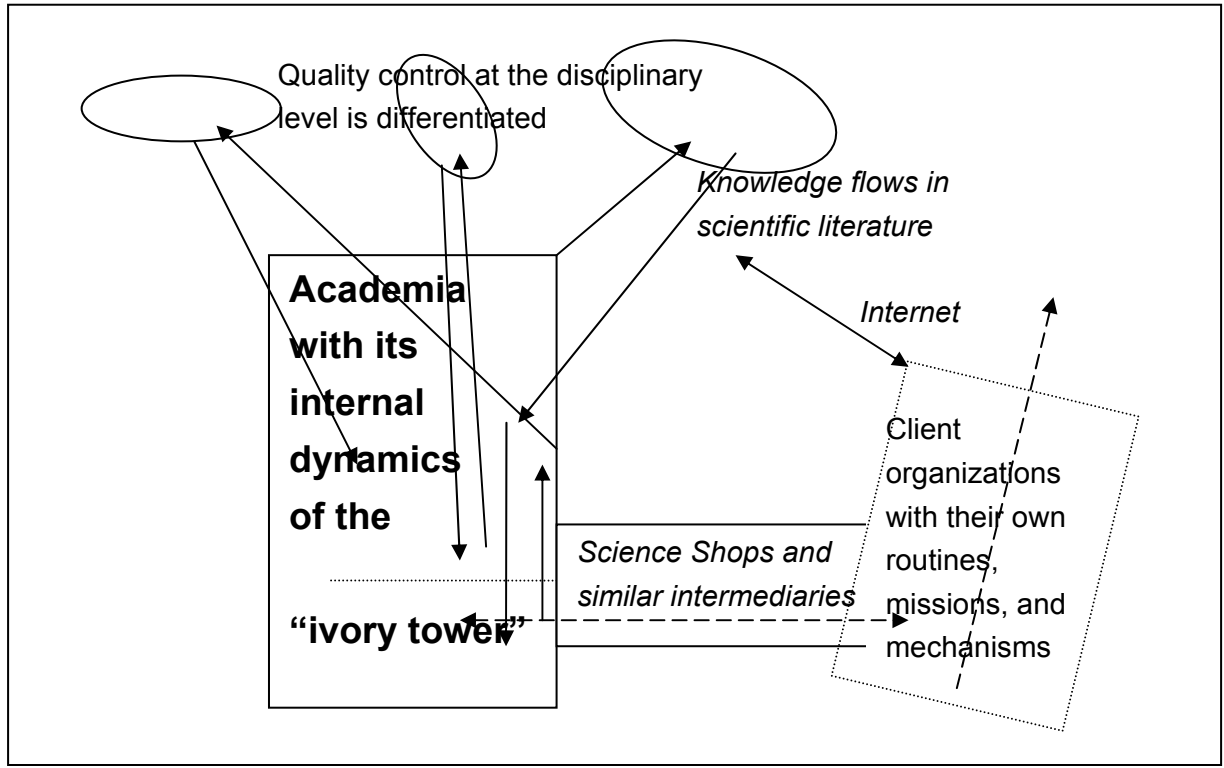

During the 1990s, the introduction and the spread of the *Internet* has changed the configuration for knowledge mediation considerably. Because communication is so central to scientific work (e.g., scientific discourse) the system itself is deeply affected by the ICT revolution (Gibbons *et al*., 1994). The external relations have been changed because increasingly, all stakeholders have direct access to relevant knowledge bases. This may vary among different groups in society, but particularly institutionalized NGOs have sometimes academics on their payroll who are perfectly able to orient themselves in relation to relevant research questions.

The new configuration is not essentially different from reading newspapers, but the costs involved in gathering information have declined with orders of magnitude. Thus, the configuration in which Science Shops can add surplus value to the information—that is, knowledge—has changed. This has implications for the question of the evaluation of Science Shop intermediations because it adds a communication dimension to these hitherto mainly local practices. In this report, we focus on this communicative dimension of the mediation process.

## 2.    Research question and methodology

The research in this subcontract to the INTERACTS project addresses the question of the external visibility of Science Shop work in terms of communications which reach beyond the local context of the participants. In addition to the question of the effects of this specific type of communication in terms of publications, institutional development, and curriculum development, we study the communication of the results in the press, the popular and grey literature, and other means of communication *insofar as retrievable on distance through the Internet*. This is an important limitation because projects and authors may be visible in other respects, which are considered more important by the Science Shops. The Internet provides us with a specific lens that enables us to provide feedback on an increasingly important aspect of the visibility and social impact of the mediation.

Because this subcontract was organized *within* the INTERACTS project as a point of external reflection, we were able to follow the reports of the individual case studies as provided by Work Package 4.[1] Each contractor provided three detailed case studies from the perspective of the local Science Shops. These 21 case studies were selected by the respective contractors as "best practices" on the basis of a number of criteria. The case studies were preceded by pilot studies in order to harmonize the methodologies (Work Package 2). In this pilot stage, the design for the final case studies could also be fine-tuned with the demands of our relatively external reflection. Thus, the case studies provided us with the basic materials as input to the analysis.

The methodology was as follows:

1.  The first full drafts of the case study reports (available as of 19 January 2003 or shortly thereafter)[2] were scrutinized for external references, names of authors, websites and other information that can be accessed from a bibliographic point of view.
2.  Each of these leads were followed-up using the *Science Citation Index* and the *Social Science Citation Index* for the scientometric evaluation, specific webpages of authors and institutions, webportals of newspapers, Amazon.com and its national derivatives (like amazon.co.uk and amazon.de), and the integrated library system of the Netherlands PICA in the case of retrieving books, as well as the *AltaVista* Advanced Search Engine for identifiable clues like the ones mentioned. The *AltaVista* Advanced Search Engine was used among the many possible search engines because of its use in other webometric research (and therefore the availability of software and routines) and its option to search for different domains in Boolean relations to specific time frames (Leydesdorff, 2001).

[1] Final versions of these reports will be made available at
http://members.chello.at/wilawien/interacts/reports.html.

[2] January 19, 2003 was the official deadline for the reports, but first versions came in until February 10, 2003.

3. On the basis of these searches a bibliographic/infographic profile is sketched for each of the 21 case studies describing the main researchers involved, the output, the scientific institutions, and the role of the Science Shop in the mediation. Conclusions in the different dimensions of potential impact like higher education, scientific publications, newspapers, etc. are elaborated and policy recommendations for enhancing the visibility of Science Shop research are made.

4. Preliminary conclusions were reported to the internal meeting of the consortium at Innsbruck (Austria), 7-9 March 2003, as a contribution to the formative evaluation. An initial version of this report was circulated among the contractors in May 2003 for comments. Note that this report is entirely based on using the first (draft) version of the full case study reports. The final versions of the case study reports (at http://members.chello.at/wilawien/interacts/reports.html) may differ from the information basis used for this evaluation.

5. In response to the initial report, it became clear that the retrieval of information about the researchers and clients in the case studies of Vienna could mean a breach of the guarantee of anonymity provided by the research team that analyzed the case studies. Although the information was retrieved at the Internet on the basis of the previously anonymized reports only, we agreed to further anonymize paragraph 3.3 in this meta-evaluation. The format of this section therefore differs in some respects.

The report is concluded with an extensive section drawing more general conclusions from this relatively external evaluation in relation to our involvement in the ongoing process of the INTERACTS project. Policy recommendations are also provided for each of the Science Shops involved.

Several of our colleagues have commented on a draft of this report that they did not recognize their activities in it sufficiently. We deliberately chose an external perspective because the internal perspectives are sufficiently covered by the case study reports of Work Package 4. If our results show a picture different from the local or even national image of the Science Shop in case, then one should consider our perspective as *additional* information. This report based on a subcontract provides a meta-evaluation on the basis of the evaluations in Work Package 4. But one should be aware that this perspective is not necessarily central to the work of the local Science Shops.[3] Yet, we hope to be able to contribute to the further development of the mediation by making sometimes critical remarks about the external visibility of this work and its results.

---

[3] In anthropology our perspective might be called "etic" as different from the internal or "emic" perspective. The "emic" perspective provides us with thick descriptions, while the "etic" one contributes by enabling us to position this perspective in relation to other (e.g., policy) discourses.

# 3 Case Study Evaluations

The analysis of the case studies follows the sequence numbers of the 8 partners in the projects. Since partners 6 and 7 of INTERACTS collaborate locally, seven reports of containing each three case studies were available. As noted, this study was based on the versions of the report available at the deadline of January 19, 2003 (or shortly thereafter).

## 3.1 Danish Case Studies

Two Science Shops are involved in the three Danish case studies. The first is the Science Shop at DTU (The Technical University of Denmark), where the involved students study engineering. Founded in 1985, this Science Shop is a part of the Department for Manufacturing Engineering and Management (IPL). Second, the Science Shop at RUC (Roskilde University Centre), established in 1988, is part of the central administration of the university.

The goal of both Science Shops is twofold: to provide an inroad into both science and research to civil society organizations, and to provide students with possibilities for including "real life" topics as part of their curricula, through the cooperation with these groups. Additionally, Science Shop DTU aims at contributing to improving the level of education and research at this university.

Science Shop DTU was established in line with the Dutch model. This means that it aims both to give scientific access to civil organizations but also aspires to make the knowledge needs of their clients result in a more permanent impact on curricula and research. The shop is involved in two types of innovative activities: one, the development of new areas of research and teaching such as urban ecology, cleaner technology, and organic food, and two, the "development from within" in already established research areas, such as when researchers from different institutes have embedded topics from the Science Shop projects into their teaching or research activities. The Science Shop of DTU then functions as a dialogue partner.

The case studies were chosen based on the criteria used by the INTERACTS project. Two cases came from the Science Shop DTU and one from the Science Shop at RUC. The three cases share their concern with topics related to the field of environmental issues, but they differ in terms of which issues are addressed, what type of scientific knowledge was requested by the clients, how the clients used the knowledge they received, and the impact it had on the actors and the society.

### 3.1.1  The Danish Cyclists Federation (DCF)

The title of this report is "What is a bicycle?  A social constructivist analysis of the possibilities of promoting the use of bicycles."  The request for this project was made by the NGO "The Danish Cyclists Federation" (DCF) to the Science Shop at DTU.  The aim of the project was to investigate how the use of bicycles can be promoted in order to make their future use more attractive.  Specifically, the goal was to analyze the motivations that bicyclists have for their form of transportation.  This knowledge resulted in recommendations to DCF on how to encourage bicycle use rather than cars.  The analysis involved a literature review and interviews with bicyclists, politicians, and traffic planners, in order to understand their perception of the bicycle as a technology.  The project took place between February 2000 and June 2000, and involved two M.Sc. engineering students in their fourth year.  No costs were involved in this research.

*Internet Visibility of the Project*
The home-page of the Science Shop, in English, can be found at http://www.its.dtu.dk/vb/eng/index.htm.  However, the report is not easily retrievable from this website.[4]  The site is organized under categories and these reports are not available under "traffic." The website appears (at the time of this evaluation, i.e., February 2003) to have not been updated since 1999.  Furthermore, the website lacks an additional search system, for example, for searching on title words.

A search on AltaVista provided varying results.  Using the name of the supervisor a search with the terms "+Morten +Elle +bicycle" provided 8 pages, but not precisely on this project. Using the names of the students "+cykel +Luxenburger" resulted in zero hits, and "+cykel +Asmussen" showed other results not related to the project.  AltaVista points to http://www.ibe.dtu.dk/medarbej/me/me_e.htm for the homepage of Morton Elle, but an investigation shows that this site no longer exists. There is a new page under construction at http://www.byg.dtu.dk/index_d.htm for architectural engineering, but this site is not further developed in English. There are no publications retrievable in English.  Searching on the Danish word "cykel" did not provide any hits on this page.  In summary, the reports are not retrievable at the Internet.

*Individual Researcher/Mediator Exposure*
The two students that performed the research are Rune Asmussen and Jan Luxenburger. The DTU supervisor is Morten Elle, and the Science Shop manager is Michael Søgaard Jørgensen.  The NGO representative is Jens E. Pedersen.  Finally, the head of the department of Manufacturing Engineering and Management is Leo Alting.

---

[4] In reaction to the first draft of this report, the Science Shop informed us that a new website is under construction at http://www.vb.dtu.dk. However, we were not able to retrieve the report from this new site.

The home page of the client organization at www.dcf.dk contains information in English and German, but no collaboration with the Science Shop is mentioned here. A footnote found in the case study led us to http://www.dcf.dk/cyklist/cyk20101/20101-01.htm and to an on-line article entitled "Leder" (in Danish). This article does not make a reference to the collaboration through the Science Shop or to the report.  However, the results of the study have been used in a follow up from the side of the client, for example, in relation to a grant from the Ministry of Traffic regarding bicycle paths.

The supervisor was not deeply involved in the project.  He has not seem to have used the results in the project for anything in relation to his research or teaching at DTU, but he was able to use the project in order to get an understanding of the students' skills and engagement, which were useful when they discussed the subject and research area for their theses. There was follow-up at the level of individual students in their Master's theses.

The students are quoted in the case study for mentioning the advantage that their reports remain available through the Science Shop library system of DTU.  However, this availability is only an archival function within the local situation.  Although the report is provided with a number, it remains irretrievable at the Internet. It is unclear whether a copy can be ordered.

A search in the *Science Citation Index* indicates that Morton Elle, who according to the report has a focus on infrastructure planning,  provides no hits.  PICA lists a publication of this author from the journal *Built Environment*, entitled "Rethinking Local Housing Policies and Energy Planning:  The Importance of Contextual Dynamics."[5]

*Conclusion*
The project itself is well done with clear results. The Science Shop DTU made marginal use of the results.  DTU did not use the project to recruit more students nor to develop follow-up projects or new ideas for future studies. The reason may be that the main supervisor in DTU, Morten Elle, does not focus on traffic planning in his research.  The project had a focus on further cooperation (process), but not on using the results (products). The Science Shop claims that the results have been constitutive for the client's actions, but the NGO has not made this visible in its publications.

### 3.1.2 Vognporten

The title of case two is "Organic food in the day centre Vognporten – with special focus on storage and local supply of fruits and vegetables."  The request for this research, which was accomplished between February to June 1996, was made by the day care centre Vognporten to the Science Shop at DTU.

---

[5] Morten Elle, Thessa Van Hoorn, Timothy Moss, Adriaan Slob, Walter Vermeulen, Jochem Van Der Waals, "Rethinking Local Housing Policies and Energy Planning: The Importance of Contextual Dynamics," *Built environment* 28 (2002), 46-56.

Vognporten is a day care centre located in the municipality of Albertslund.  It is a pre-school institution for about 50 children, ages zero to six.  The aim of the project was to investigate the possibilities of storage and local supply of organic fruits and vegetables at Vognporten. A second objective was to establish contacts with a farmer who could provide a visiting place for the centre as well as a place to buy fruits and vegetables locally.

Vognporten originally contacted DTU in 1995 for assistance with research regarding environmental management in the day care centre.  The research was carried out by two students midway through their engineering studies, and was accomplished through a literature review and through informal talks with the day care centre staff.  It also involved a trip to Sweden to look at some igloos to use for vegetable and fruit storage, as the research showed that the most sustainable way to store fruits and vegetables is in an igloo.  No costs were involved in the research except the travel costs for the visit to the igloos in Sweden.

*Internet Visibility of the Project*
An abstract of this project could be retrieved from the website of the Science Shop at DTU at http://www.its.dtu.dk/vb/fproj/abstract/foedevare.htm, in Danish, and a less extensive English version at http://www.its.dtu.dk/vb/eng/projects.htm. The abstracts mention two pamphlets and an article in the Danish magazine *Grøn hverdag*.  This magazine has a homepage at http://www.gronnefamilier.dk/hverdag.asp , but the article cannot be retrieved from this site.

At http://www.praktiskoekologi.dk/artikler/artfrugt/opbevaring.htm we found an article by these two authors from 1997 published in *Praktisk Økologi*. The English web site of this Danish Society of Practical Ecology is located at
http://home4.inet.tele.dk/redak-km/engelsk/index.html.   The site mentions that the articles are available in Danish. The Society is part of the Danish EcoWeb at http://www.ecoweb.dk/index 2.htm.

There is also follow-up within the client organization at http://www.praktiskoekologi. dk/artikler/artbyggeri/bygjord k.htm. Here, a further explanation is given as to how one can construct the igloo as advised in these reports. References to the reports are provided.

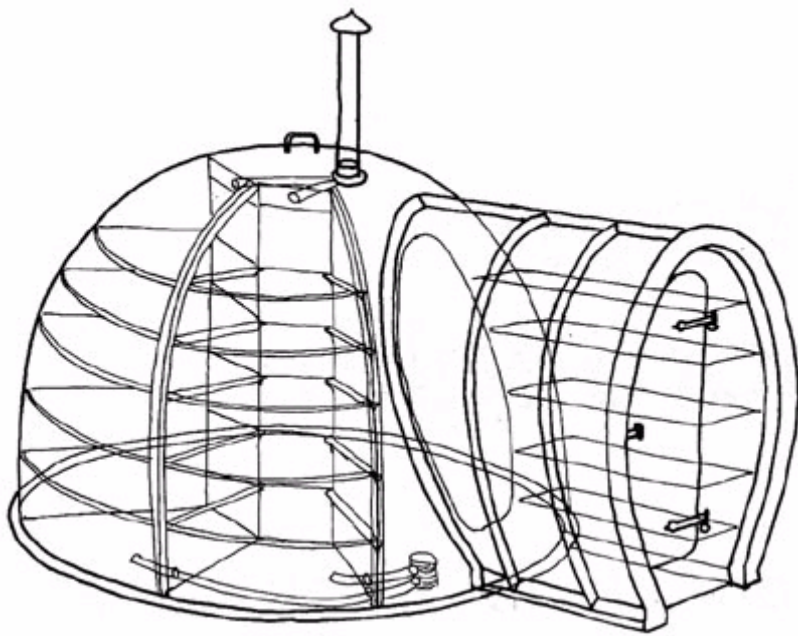

The results of this project are among those further elaborated in a book entitled *Organic Food in Catering – The Nordic Perspective,* published in March 2002. This book can be found at http://ecocater.net/media/nif17.doc and it was prepared by Bent Egberg Mikkelsen, Niels Heine Kristensen, and Thorkild Nielsen, the supervisor of the project here under discussion.  The report is based on literature, interviews, as well as papers written by Carin Enfors, Gunnar Á. Gunnarsson, Níels S. Olgeirsson, Sampsa Heinonen, Irma Kärkkäinen, and Erik Evenrud.  Two hundred hardcopies were made available by the Danish Veterinary and Food Administration, for the price of DKK 66.00 each.

A search on AltaVista for "+Vognporten +Albertlund +DTU" returned 6 hits, but none of which were relevant to the project.  Searching for Susie Barfod (later name of Susie Ebbesen) resulted in 12 hits, again not related.

*Individual Researcher/Mediator Exposure*
The students's supervisor was Thorkild Nielson, and his main research focus is research in relation to organic food.  The Science Shop manager is also in this case Michael Søgaard Jørgensen.  One of the researchers for the Eco Group is Niels Heine Kristensen.

The student report, "Økologiske fødevarer i daginstitutionen Vognporten – med fokus på opbevaring og lokalforsyning af frugt og grønt," resulted in two hits from a search on AltaVista.  One of these sites can be found at
 *http://www.praktiskoekologi.dk/artikler/artbyggeri/bygjordk3.htm*.  As mentioned in the report, these two students further pursued their interest in soil problems towards the Master's thesis.  At *www.isva.dtu.dk/grc/pdf/GRC-99.pdf* a copy of the 1999 annual report of the Groundwater Research Centre at the Technical University of Denmark can be retrieved.

The research done by the students has further stimulated the so-called Eco-group in recognising that student projects through the Science Shop, or without the involvement of the Science Shop, can be relevant for the research done within this researchgroup.  Student projects often contribute to the development of new ideas and perspectives related to the research area.

The supervisor has a homepage at http://www.its.dtu.dk/ansat/tn/tn_e.htm, which was last updated in August 2000.  Three recent articles were found on this website with Thorkild Nielsen and Niels Heine Kristensen as authors.  Additionally, Nielsen is part of the Eco Research Team at the Department of Technology and Social Sciences, Technical University of Denmark.[6]  This information can be found at http://www.its.dtu.dk/proj/okopro/index.htm. The team is also part of the Eco-Web that was encountered in the first case.  At this site a list of publications of Niels Heine Kristensen and Thorkild Nielsen (1992-1998) can be retrieved, mainly in Danish.  One paper is listed in English, "From Social Movement to Food

---

[6] The name of the department was changed in January 2001 into "Department of Manufacturing Engineering and Management. However, the publications of this researcher are still located at the old address (11 July 2003).

Industry. Social Actors, Strategies and Regulation of Organic Food Production." These publications can be ordered and the webpage is maintained, but there is no publication activity visible since 2000. (The last update was on August 8, 2000).

The supervisor has further used both the results of the research and the institution as an illustration in different seminars arranged by the Eco-group. A search on AltaVista let to this link at http://www.jrc.es/iptsreport/vol20/english/FOO3E206.htm, which has an article in IPTS Report No. 20 entitled "From Alternative Agriculture to the Food Industry: The Need for Changes in Food Policy," The article was published in December 1997 by Niels Heine Kristensen and Thorkild Nielsen at the Technical University of Denmark.
An additional search for Laila Rasmussen +Vognporten retrieved no relevant hits.

A search for "T Nielsen" and Denmark provided 58 hits in the *Science Citation Index*. "T Nielsen" and Lyngby: 2 hits, but both were located at a fishery institute (not from this author).

*Conclusion*
The research showed that the most environmental sustainable solution to storage of vegetables and fruits was underground storage in an igloo. As a direct result of the research, Vognporten was able to get funds from the municipality in order to buy such an igloo. No follow-up was found on the side of the Science Shop, but there was substantive communication to the relevant audiences. The Science Shop claims to have been constitutive to this research program during the 1990s, but this is not made visible by the researchers involved at the level of their department.

### 3.1.3  DN Frederikssund

The project's title in English is "Biomanipulation in shallow eutrophic lakes – a study of food web interactions and lake equilibria". It was written by Tine Amhild, Jill Grenaae, Søren Olsen, and Louise A. Zimmer. The request for the research was made by a local committee of the NGO, the Danish Society for the Conservation of Nature (DN) in Frederikssund, through the Science Shop at Roskilde University Centre (RUC).

The research questions were twofold: one, is it possible, through biomanpulation, to reach a sustainable situation resulting in the water being clear in an eutrophic lake? Two, is this situation possible in the Lille Rørbaek village pond? The research consisted of theoretical considerations regarding biomanipulation in shallow eutrophic lakes, as well as tests and water samples done in the Litte Rørbaek village pond. The students consisted of four Environmental Biology students in their fourth semester, and the research took place between February to June 2001. The costs were covered by the students' institute.

*Internet Visibility of the Project*
The Science Shop Roskilde can be found (in Danish) at http://viden.ruc.dk/. The titles of the reports are provided,[7] but they cannot be accessed at the Internet.

*Individual Researcher/Mediator Exposure*
The homepage of the researcher and supervisor involved, Benni Winding Hansen, can be retrieved at http://virgil.ruc.dk/~bhansen/. The web-page is kept up to date. Hansen is a lecturer at the Environmental Biology institute at RUC, and his main focus is on aquatic marine biology. The website of the local committee of DN Frederikssund at http://www.home7.inet.tele.dk/hbc is no longer retrievable. The students are not visible on the Internet. The researcher (Hansen) is a prolific author, very visible with several publications per year in the *Science Citation Index*.

*Conclusion*
The project consisted of junior students with high-quality supervision on the basis of a request by a large environmental organization. The project resulted in the supervisor having hired one of the students as an assistant and having recruited two of the students to do their Master's theses, in connection with one of his research projects. Additionally, this project contains very positive evaluations of the higher-education component because undergraduate students took the initiative for organizing this project.

### 3.1.4 Conclusions and Recommendations

The Danish Science Shops have an established practice of mediation to the higher-education system. Students who participate, seem to qualify for assistanceships. Because of this successful practice, these Science Shops have hitherto perhaps payed not sufficient attention to the research function and the communication of results. The Science Shop at DTU has as one of its objectives to initiate new research lines and it seems to have been successful in that, but this is not made visible from the outside. The researchers do not acknowledge its role and do not refer back to the Science Shop.

Although the external relations seem to function well, the clients are not inclined to make reference to Science Shops or the university in their publications based on project results.

---

[7] Danish report title: "Biomanipulation I lavvandede, eutrofe søer – et studie af interaktioner I fødenettet og ligevægtstilstande" Udarbejdet af Tine Amhild, Jill Grenaae, Søren Olsen og Louise Aa. Zimmer. *Videnskabsbutik* nr: 97.43.

English report title: "Biomanipulation in shallow eutrophic lakes – a study of food web interactions and lake equilibria". Written by Tine Amhild, Jill Grenaae, Søren Olsen og Louise Aa. Zimmer. *Videnskabsbutik* nr: 97.43.

Recommendations:

- Improve on the existing databases of the Science Shops at the Internet. Make these databases searchable on authors and subjects without imposing your own (ex ante) categories as the single search terms.
- Relate to the researchers involved as supervisors by hyperlinking to their home pages and provide electronic photographs of both the people involved and the situations under study. Make the existing network relations externally visible by generating more acknowledgements, references, and hyperlinks.

## 3.2    Austrian Case Studies I: Tyrol

Three mediating institutions were involved in the case studies for Innsbruck.  First, the Institute for "Forschung, Bildung und Information" FBI is a non-university based Science Shop and research institution, in operation since 1991.  Its main goal is to bridge the gap between the university and the public in making advanced knowledge accessible, understandable and applicable for a broad public.  It serves as a link between academia and society as well as between theory and practice, on issues related to research, society and culture, with a special focus on women and gender issues.

Second, the Wissenschaftsagentur Salzburg is a university-based Science Shop located in Salzburg.  It is organized as a non-profit organization and works to provide knowledge transfer between the University of Salzburg and society at large.  Third, Patenschaftsmodell Innsbruck (PINN) is a Science Shop equivalent, specifically a service center for enterprises and organizations, at the Faculty of Social and Economic Sciences at the University of Innsbruck. PINN aims at building up contacts between university and practice, on a systematic and regular basis.  Additionally, it is aimed at promoting the practice orientation in the higher education offered by the Faculty.

### 3.2.1  Customer Satisfaction of the Aggrieved in Mediation in Penal Matters

The project under study evaluated customer satisfaction in relation to a service called "mediation in penal matters," which is provided by the NGO.  The NGO, the association for probation services and social work (ATA), is a small local branch of a large, nationwide NGO.  It offered services with the objective of bringing about reconciliation between victims and suspects through mediation.

The NGO approached PINN based on their conviction that it is important to emphasize that client's satisfaction is a timely requirement, which is important not only for businesses but also for NGOs working in the social field.  The evaluation was conducted by two final-year undergraduate students in the Faculty of Social and Economic Sciences.  The main research questions involved such areas as the organizational boundary conditions, the attitudes of the staff, the process itself and the assessment of the results, all aimed at improving the level of service offered.  The methods includes a survey, semi-structured interviews, and a moderated group discussion.  The duration of the project was from October 2000 until May 2001, and the costs were € 1090.

*Internet Visibility of the Project*
Patentschaftsmodell Innsbruck can be found at
 http://info.uibk.ac.at/c/cb/cb19/haupt_e7.html.  It is part of the Sozial und Wirtschaft-wissenschaftliche Fakultät of the Leopold-Franszens-Universitaet Innsbruck, and is a transfer unit which mediates with third partners.  It provides service for both business and organizations.

A primary publication from this project can be found in SUB, a professional journal: Andrea Altweger and Evelyn Hitzl, "Außergerichtlicher Tatausgleich - Kundenzufriedenheitsanlyse der Geschädigten," *SUB* 01/2002, 24. Jahrgang, S.17-22, Wien. This publication cannot be retrieved on the Internet. Furthermore, a review is mentioned in TOA Infodienst, a professional journal: Bernhard Trummer-Kaufmann, "Zufriedene Geschädigte im Außergerichtlichen Tatausgleich," *TOA Infodienst, Rundbrief zum Täter-Opferausgleich*, Nr. 16, April 2002, S. 33-36, Köln. This report could not be found, but a similar study is located at http://www.iuscrim.mpg.de/forsch/krim/kilchling2_e.html. This article deals with a comparison between German and Austrian practices, part of an ongoing debate in Germany and Austria. Additionally, literature surveys are available on the Internet.

At http://www.dbh-online.de/html/publikationen.html, both TOA–Infodienst and Rundbrief zum Täter-Opfer-Ausgleich are mentioned. An article in the *Salzburger Nachrichten*, a regional daily newspaper, is available at
http://www.salzburg.com/servlet/scom2/searchresult?xm=199011&res=0. The authors are not mentioned, but a reference is made to the university.

The homepage of the NGO could be found at http://www.neustart.at/angebote/ata2.php.
It has available two brochures (one in German and one in English) about ATA, but there is no specific mention of this research project.

*Individual Researcher/Mediator Exposure*
A Master's thesis written by Andrea Altweger and Evelyn Hitzl, resulted from this project. It is entitled "Kundenzufriedenheitsanlyse der Geschädigten im Außergerichtlichen Tatausgleich." The thesis cannot be retrieved among the "Diplomarbeiten" at the website of the PINN (the search +Hitzl +Altweger returned zero hits). The supervisor is not visible on the Internet.

The organizer of the project was Professor Stefan Laske, head of the department. Information on him and his report ("EUROMOBIL - Ein praxisorientiertes Förderungsprogramm für leistungsorientierte Studierende der Wirtschaftswissenschaft") is located at http://iol1.uibk.ac.at/cgi-bin/iol_mitarbeiter_detail.pl?nr=12. The supervisor of the project, Martin Piber at the Department of Organisation and Learning, can be located at http://iol1.uibk.ac.at/cgi-bin/iol_mitarbeiter_detail.pl?nr=4. No other publications are sited in this line of research. There is also no retrieval of publications by these authors by using international databases.

*Conclusion*
The project is typical for a PINN project, in that students carry out the project within the context of their Master's thesis under supervision of academic staff. It was not typical that the client organization was an NGO. The customer satisfaction was high, and one was appreciative of the service offered by the university. The external evaluation by these students provided insights in aspects where improvements could be made, such as in the sensitivity and competence of staff members, and in helping the victim to gain an understanding and appreciation for the fact that the suspect had taken on the responsibility for his crime.

### 3.2.2  Children and Young People in Lungau:
### Between Participation and Apathy

The request for the research was made by the managing director of the NGO to the Wissenschaftsagentur Salzburg.  The NGO, Akzente Salzburg, is a medium-sized NGO in Salzburg, working with youth in offering a wide range of activities.  This project investigated the quality and circumstances of life for young people in Lungau, a rural region of the federal state (Land) of Salzburg.  It specifically aimed at finding out their needs and desires.  The study was intended to provide a well-grounded scientific basis for the stimulation of youth work, for the establishment of a youth center and youth information point, and furthermore for initiating future youth projects in this region.

The study was conducted by one final year undergraduate student, in the context of a Master's thesis.  The main research question focused on leisure behaviour and leisure amenities, labour supply, labour market and working conditions, the effects of tourism, participation in a political context and standards and moral concepts.  The duration of the project was between January and September 2001, and the costs were € 6500.

*Internet Visibility of the Project*
Wissenschaftsagentur Salzburg is located at http://www.sbg.ac.at/was/.  A report about the project is made in the *Zeitschrift der Universität Salzburg* at
http://www.sbg.ac.at/plus/plus_3_2002/rubriken/wissenschaftstransfer.htm.  The project is acknowledged as a product of the Wissenschaftagentur of the University of Salzburg.  The graduation of the student involved is documented at
http://www.sbg.ac.at/aktuelles/presseinformationen/2002-01-16.htm.

AltaVista listed two results, located at
www.land-sbg.gv.at/regierung/burgstaller...es/pa1129w2.htm and
www.land-sbg.gv.at/lkorr/2001/11/29/26635.html.  Additional information is provided at the page of Akzent at http://www.akzente.net/.  A newspaper article is mentioned at http://www.lungau-aktuell.at/aktuell/aktuell.htm, but the report cannot be retreived from here.

*Individual Researcher/Mediator Exposure*
The Master's thesis which resulted from this project, "Kinder und Jugendliche im Lungau zwischen Partizipation und Apathie", by Neubacher Dagmar, Dipl. Arbeit Salzburg 2001, cannot be retrieved under the reports which can be ordered on-line at the site of the PINN office. The client, Akzente Salzburg, can be found located at http://www.akzente.net/, and the client group at
http://www.salzburg.gv.at/themen/politik/landesregierung/burgstaller/schwerpunkte/jugend_.htm.  But there is neither mention of the report nor of the collaboration.

The home page of the supervisor (Univ.-Prof. Dr. Herbert Dachs) is found at http://www.sbg.ac.at/pol/people/dachs/dachs-pfad.htm. The supervisor is an established researcher with an impressive publication list. The study is not central to his research interest.

*Conclusion*
As intended, a sound scientific basis was generated to continue youth work in the region and help to develop a tailor-made package of measures as well as provide opportunities for the young people.  The report is not retrievable and the results cannot be found in the archival records of the Internet. However, two brochures have been distributed widely at the regional level.

### 3.2.3  Precaution against heart disease for Turkish migrant women in Tyrol

A request for the research was made by the NGO to the Science Shop, Institute FBI.  The project evaluated a series of lectures involving precautions against heart disease for Turkish migrant women in Tyrol. The lecture series was conducted twice (1999 and 2001) by the NGO, Ludwig Boltzmann Institut für Kardiologische Geschlechterforschung.  This NGO is located in the health sector with a focus on women and the aim to support women.

Specifically, the research aimed at discovering reasons for the decline of the number of participants in the second round of lectures in 2001, and additionally to make suggestions on how to reach the target group more directly and effectively, in including a more general spectrum of Turkish migrant women.  This evaluation was conducted by two researchers who were staff members at the Institute FBI, as well as two medical students with Turkish origin, who worked as interpreters and experts of the cultural background.  The project took place between September and December of 2001, and the costs were € 3270.

*Internet Visibility of the Project*
A search using the name of one of the two researchers "Annemarie Schweighofer" received 399 hits on AltaVista (February 2003), and the other researcher's name "Gabriela Schroffenegger" had 40 hits.  The combination with AND provided 14 hits. Some other work of these authors was mentioned under the heading of "Geschlechterforschung" (gender studies), but never specifically related to this report.

A book was found on amazon.de, entitled 'Eigentlich lief alles nach Plan, bis . . .,' with these two authors as well as two additional authors. Both these authors are very active in publications, and one of them is also visible as participating in the social-democratic party.

The client is found at http://www.ludwigboltzmann.at/institute/institute_info.php?a_id=134 and http://lbi-frauen.uibk.ac.at/deutsch/.  The report cannot be retrieved, as the Insitute seems to be small, but it is very well networked. The central figure is Univ. Prof. Dr. med.

Margarethe Hochleitner (who is also the interviewee number one). No hits were located for this researcher's name in the *Social Science Citation Index* or *A&HI*.

*Conclusion*
The project was rather typical for Institute FBI, with respect to the methods applied and the intensity of cooperation. It is not typical, however, in that there was a lack of public relations and a limited dissemination of results, which were restricted exclusively to internal usage by the NGO. The researchers of the Science Shop itself are otherwise prolific authors. Some key issues that emerged from the interviews are the importance of an independent, external expert for project evaluation, the importance of qualitative methods and approaches in the evaluation, and the recognition of an expert role of the fringe group.

### 3.2.4 Conclusions and Recommendations

Institute FBI (Forschung, Bildung und Information) Innsbruck is the name of the Science Shop that is a partner in INTERACTS. Its homepage is at http://www.uibk.ac.at/c115/c11508/index.html, but this address is not even mentioned in the case study reports written by this Science Shop.[8] Although there are several publications available from the noted site, the three projects discussed below are poorly documented, including the third project which was done by the FBI itself. Of course, there may be privacy reasons and there is always an issue of ownership of the reports, but the global dissemination strategy in the three cases discussed in this report is not sufficiently developed, in our opinion.

- Perhaps, the existing database can be extended with all available reports and links to science shops and other mediating institutions in the region and the reports of these agencies can be actively included wherever possible.
- Academics with and for whom the Science Shop works can be noted to incorporate mentioning of this origin in their presentations and publications. Perhaps, it is an idea to offer co-authorship. (The FBI institute obviously has the competences in house).
- Remain focused on quality using the standards of FBI since these authors are prolific and successful. The work of FBI provides a specific focus and expertise, it is published in local and national media, but it is hitherto not sufficiently accessible at the level of the Internet or the international literature. Given the European dimension of the INTERACTS project, this additional dimension may become more important.

---

[8] This omission in the draft version was corrected in the final version.

## 3.3 Austrian Case Studies II: Vienna / Graz

Two cases were chosen from the Science Shop Graz and one from the Science Shop Vienna. All three cases represent cooperation of NGOs with universities, researchers, and students. One of the cases involves a larger project with several collaborating student researchers, and the other two involved only one or two MA students.

As noted in the methodological chapter there are severe limitation on reporting on these cases because the researchers and students have explicitly wished to remain anonymous. Thus, we don't provide names or web pages which can lead to their identification. The description in this section may remain a bit vague since we study here the visibility of reports which we were not allowed to mention.

### 3.3.1 An Empowerment Project in Vienna

This case involved an NGO focusing on empowerment in a neighbourhood. The Science Shop Vienna provided the mediation and performed the scientific evaluation of the project. Students and supervisors came from the University of Vienna (Institute of Sociology), the Vienna University of Economics and Business Administration, and the Vienna University of Life Sciences.

The project was initiated by a social centre in the neighbourhood. The aim was to improve the lives of tenants in this settlement by giving them more responsibility and autonomy. Research methods included interviews with inhabitants, field analysis, statistical analysis, analysis of media, and questionnaires. The project's duration was June 1995 until August 1998. There was no budget except reimbursement for expenses.

*Internet Visibility of the Project*
The study can be retrieved at several webpages (but, as noted, for reasons of anonymity these sites cannot mentioned here). A detailed description of the project can be found at the homepage of the main researcher.

A search in AltaVista using the name of the project provided 22 hits. From one of the results, a link with an incorrect address is provided back to the Science Shop. At this location, the university is not mentioned under the heading of "involved persons." The description provided mentions that there has been a subsidy of the City of Vienna. As noted, the Science Shop project is completed without payment (except reimbursement of expenses). The AltaVista search also supplies a non-functioning link to the website of the NGO. A critique on the project and the roles of the NGO could also be found.

A book regarding the project, but with different authors, could be retrieved at the ZSI Bibliotek. The book was published in Vienna in 1999. The book title cannot be retrieved using amazon.de, and also not in Pica. There are perhaps contributions by the authors of the various MA thesis, but they nor the Science Shop or the university are made visible at this address.

Another link provides an overview of the site PlanSinn concerning this project. This project is co-financed by the city government and one of the ministries (BMUJF). The collaboration with the university is not mentioned, but this took place in the period 1996-1997.  Perhaps, the results of the project were not yet available at this time. Other collaborations with the university, however, are mentioned.

*Individual Researcher/Mediator Exposure*
Because of this project, six masters theses were written, three of these supervised by a university professor at the Institute of Sociology of the University of Vienna.  The Theses are available for the price of  € 36 ( + € 4 porti) at the Science Shop.

One of the student's homepage could be retrieved. Here, three publications are listed as well as four presentations.  An AltaVista search using the student's name returned 27 hits. Among these, relations with the Science Shop are made visible. A summary of the thesis is available at different places on the Internet.

A publication list of the supervisor is available at a page belonging to the site of the Sociological Institute of University of Vienna. The supervisor is a prolific author with a series of monographs, book chapters, articles, etc. Two publications could also be retrieved from the (international) *Social Science Citation Index*, but only the second one is probably to be attributed to this author.

*Conclusion*
In summary, case one has provided a well-established research line. The project is part of a series of projects developed by an active researcher of this institute. The MA student whose thesis was central to the case study report, was hired by another department of the university.

### 3.3.2  Volunteers and buddies for mentally disordered persons

As mentally affected persons are always in danger of becoming completely isolated, it is important to find volunteers to provide companionship and to improve the quality of life. Specifically, this NGO wanted to know if such social companionship is successful.  The project focused on the needs and desires of the "buddies," and involved a theoretical analysis of volunteering and qualitative interviews with 14 people.  The project's intermediary was the Science Shop Graz, and the student and supervisor came from the University of Graz. The project lasted from spring 2001 to 2002, and there was no budget involved. The NGO, which is financed by public funds, works with volunteers to provide care for persons with mental diseases or disorders.

*Internet Visibility of the Project*

The website of the Science Shop in Graz is located at http://www-gewi.kfunigraz.ac.at/wila/index.html. There is a listing of theses, but no authors or links to the departments are provided. An overview of publications is also provided. Prices are more moderate than those in Vienna, around € 9. Unfortunately, the project discussed in this report does not seem to be among them, nor is the specific thesis available that was generated from this project. The graduation date of the student-researcher is mentioned by the Science Shop. The title of the scientific report of this project (2002) is also provided at this place, but no further information is retrievable.

*Individual Researcher/Mediator Exposure*

The project is supervised by a university professor. The (anonymous) supervisor has an impressive list of publications. Most of the titles are in German. Nothing further was found using AltaVista for a search with his name. A search for publications in the *Social Science Citation Index* returned no results. In PICA, two publications were located, one in 1987 and one in 1988.

It is mentioned in the case-study report that the main role of the NGO in this project was to provide the interviewees. Otherwise, the project was developed using the academic standards of the specialty.

*Conclusion*

This seems a standard MA Thesis with professional supervision, but on the basis of an interaction with an external client. It is not so clear from the information provided by the various partners at the Internet what happened with the results other than that the student graduated.

### 3.3.3  Child Poverty in Austria

The NGO was interested in receiving sound scientific information on government subsidies for families in the course of its daily activities and directed a request to the Science Shop Graz. The inquiry was made in the spring of 1998, the first Master's Thesis was completed in the winter of 2000, the second in the spring of 2001, and a press conference presenting these results took place in the fall of 2001. There was no budget in the planning of this project.

The NGO does no longer exist. Its aim was to change the general social conditions for the establishment of a children-orientated society. It dealt with issues such as children poverty, subsidies for families, the safeguarding of children, etc.

*Internet Visibility of the Project*

The two projects are described in a webpage of an Internet information center at http://www.korso.at  The two cases are not made visible among the publications of the Science Shop in Graz at http://www-gewi.kfunigraz.ac.at/wila/publi.html.  However, one project (2001) is listed among the list of finished projects. The second MA Thesis is not listed.

A search on Google retrieves a website of the department which lists a complete overview. In the archive of the *Kinder und Jugend Anwaltschaften* (Children and Youth Lawyer Organizations) an article by the two MA students involved could be retrieved.

*Individual Researcher/Mediator Exposure*
The graduation of one of the MA students in 2001 is mentioned at the website of the department.  However, the supervisor remains invisible at the site. This university professor commands an impressive list of publications. One co-authored publication between the student and the professor is among the results of the project.

Another website of the university includes information and a photo of the second student. She was subsequently hired by the department.  A search for the supervisor in the *Social Science Citation Index* returned two hits, one in 1992 and one in 2000.  The students received no hits hitherto in the *Social Science Citation Index*.

*Conclusion*
From an academic perspective, this project was very successful.  One of the students was hired by the department, and the other has coauthored a publication with the supervisor.

### 3.3.4  Conclusion and Recommendations

The first general observation about these case studies is that the Science Shops have perhaps not organized the outcome of the projects satisfactorily for serving their own ends. The academic standards of these projects are high, but the MA Theses are not published by the departments themselves (because MA Theses are considered as grey literature). However, two of these students obtained academic jobs with follow-up of the projects in publications. Another student has a publication coauthored with the supervisor. These are impressive results. All these projects were well-embedded and supervised in the respective university departments.

It transpires from the INTERACTS case study report that there have been several conflicts of interests in the mediation, but the academic side has been successful in handling these conflicts. The Science Shops involved may wish to reflect on their own role and follow-up. These projects are excellent representatives of Science Shop mediation, but they may not have been fully exploited from this perspective. The Science Shop can perhaps find more academic recognition for the quality of both the research process and the mediation.

The legal limitations on publicizing academic results because of the wish of researchers and students involved to remain anonymous, in our opinion, indicate perhaps an over-politicization of some of the issues involved. The limitations placed upon our reporting in this section may serve as an example.

## 3.4    German Case Studies

The Berlin's case studies come from two different Science Shops.  The first, *kubus*, the Kooperations und Beratungsstelle für Umweltfragen (Co-operation and Consulting for Environmental Questions), is a Science Shop located at the Technical University Berlin.  The second, Science Shop Bonn (Wissenschftsladen Bonn), focuses on ecology and environmental protection.  Founded in May of 1984, it works to make scientific information and results accessible to institutions, groups, and occasionally individuals who do not have access to such results.  WiLa Bonn has contacts with scientific experts all over Germany but is not affiliated with the University of Bonn.  The staff of this Science Shop hold a great deal of expertise in this field and therefore are well established in the region.  Services include such activities as counselling, project-development, workshops and conferences, reports, surveys, and newsletters.

Both German Science Shops, *kubus* and WiLa Bonn, are retrievable on the Internet.  The homepage for *kubus* is located at http://www.tu-berlin.de/zek/*kubus*.  Searching with "*kubus*+berlin+tu" led to 388 hits.  For WiLa Bonn, the homepage can be found at http://www.wilabonn.de/.  A search with "+wila+bonn" provided 523 hits.  Both of these institutions are professionally based and therefore do not include the involvement of students.

### 3.4.1  Tiergarten Tunnel

As a part of German reunification, many construction projects have been planned and carried out in the capital city of Berlin.  One of the largest of these projects is a system of tunnels for railway and motorway use under the largest public park in Berlin, the Tiergarten.  The project is known as the "Tiergarten Tunnel."  Since the project has generated a great deal of controversy from different action groups because of social and environmental impacts, an umbrella organization was formed in order to address these objections.  This organization was called the "Anti-Tunnel GmbH" (ATG).

A German "Federal Nature Conservation Law" states that NGOs, which are recognized by the administration, are allowed and requested to report on concerns about construction plans that could detrimentally affect the environment, and these statements will be considered by the appropriate departments of the administration.  Further, recent developments have allowed certain NGOs to bring action against governmental institutions and the administration if evidence demonstrates that construction measures are causing significant damage to the environment.

In 1994 ATG first contacted *kubus* for the formulation of an expert-report, or environmental impact study, in their legal case against the tunnel project.  The expert-report was needed because the politically motivated top-down decision making in Berlin encouraged the continuation of the Tiergarten tunnel. Additionally, the influence of the construction lobby was felt. The counter charges were presented in court.

The general scope of the report answers the question, "What are the environmental risks of constructing the tunnel in the Tiergarten area?" This question was answered in detail in several ways. First, expert reports were formulated concerning the ecological effects of the tunnel. Second, new studies were completed in order to supplement the analysis and alternative recommendations were made. The help and resources of *kubus* were requested because the ATG was not able to financially support private research institutes, and the scope of questions was so large that it was felt impossible to operate on a voluntary basis. With *kubus* acting as moderator, an effort was made to build an anti-publicity campaign as well as a scientific anti-initiative in response to the tunnel plans.

The cooperation between ATG and *kubus* ended with the court case, which took place in July 1994. It was decided that evidence in the counter-reports for the environmental risk of such an endeavor was not sufficient to stop the construction of the tunnel. The tunnel is currently being built and will most likely exceed its predicted costs.

The research was financed with about € 13,000 from the Stiftung Naturschutz Berlin (Berlin Foundation for Environmental Preservation). *Kubus* did not receive additional funding, and the staff and infrastructure were financed by the Technical University Berlin.

*Internet Visibility of the Project*
The case study report mentions two accompanying studies published during the construction measures on behalf of the "Berliner Landesarbeitsgemeinschaft Naturschutz" (BLN - Berlin Working Group on Nature Conservation). One of these was funded by Stiftung Naturschutz Berlin and entitled "Baumvitalität im Berliner Tiergarten während der Großbaumaßnahmen;" the other by the Berlin Authority for the Preservation of Monuments and entitled "Baumvitalität im Berliner Tiergarten." Both reports produced from this case can be ordered from the website by email. However, the reports are not available in .pdf format as they were produced in 1995.

The *kubus* website also mentions the availability of a book available for DM 50, entitled "Baumvitalität, Wasserstatus und Altlastenproblematik im Bereich des Tiergartens."[9] Additionally, the bibliography of the case study report list 15 publications, beginning in 1979, that are related to the Tiergarten Tunnel project. These publications reflect both a focus on the project and the aftermath of the project. The Tiergarten Tunnel project fits into a practice and a potential market for expertise, for example at http://www.stadtenwicklung.berlin.de/umwelt/monitoring/en/vegetation/. On its completion, the project received a great deal of attention on the Internet, but this exposure was not specifically related to this project (at

[9] Barsig, M., Bisom, N., Wichmann, A., Paust-Lassen, P. (1995): „Baumvitalität, Wasserstatus und Altlastenproblematik im Bereich des Tiergartens". Berlin: BLN in Kooperation with *kubus*, 158 pp.

http://morgenweb.de/archiv/2001/05/14/aus_aller_welt/20010514_15_RC01237023_13301.h
tml).

*Individual Researcher/Mediator Exposure*
The main researcher could be identified as Michael Barsig, and is a researcher at the
Institute of Biology and Ecology, TU Berlin.  The main mediator is Wolfgang Endler, a
researcher at the Science Shop and also responsible for this project. Through an Internet
search, Michael Barsig could also be retrieved in relation to the advisory practice of Nicolas
A. Klön on tree-diagnostics in Berlin, located at
http://www.baumdiagnostik.de/index.html.

The long-term collaboration between Barsig and Endler is demonstrated in the inclusion in
the *Science Citation Index* of two international publications that they co-authored in
*Angewandte Botanik* in 1990[10] based on a 1989 report on the subject in German.[11]  Barsig's
dissertation can be purchased on the amazon.de website.  Currently, Michael Barsig seems
increasingly to have chosen to pursue a scientific career, and Wolfgang Endler has become
a staff member at the Science Shop *kubus*.

Two more recent, international publications of Michael Barsig could be found using the
*Science Citation Index*.  These publications, in 1998 and 2000, were internationally co-
authored with Swedish authors, and are also retrievable through PICA. Barsig's PhD Thesis
(2002) on the subject can also be retrieved within German databases, but not abroad (e.g.,
in the Dutch *PICA*).[12] The Thesis, however, is mentioned on the pages of the city planning
agency  at  http://www.stadtentwicklung.berlin.de/umwelt/monitoring/en/vegetation. There  is
an obvious interest among larger audiences in the results of this research.

*Conclusion*
The Tiergarten Tunnel project is strongly paired with the environment in Berlin, but it is
overshadowed by a great deal of other activities and actions.  For example, two students of

---

[10] Endler W, Barsig M, Weese G, Hafner L  (1990). Cytomorphological Investigations of Pine Needles
(Pinus-Silvestris L) in a Conurbation .1. The Ultrastructure of Chloroplasts in Mesophyll According to
Season, Age, and Pollution Burden, *Angewandte Botanik*, Vol 64, Iss 3-4, pp 289-302;

Barsig M, Endler W, Weese G, Hafner L  (1990). "Cytomorphological Investigations of Pine Needles
(Pinus-Silvestris L) in a Conurbation .2. The Spectrum of Ultrastructural Damage Phenomena,
*Angewandte Botanik* 1990, Vol 64, Iss 3-4, pp 303-315.

[11] Hafner L, Endler W., Wendering R., Barsig M., Freitag G., Klein G. & Weese G. (1989).
"Feinstruktur  der  geschädigten  Kiefernnadel.  Abschlußbericht  im  Rahmen  des  FE-Vorhabens
"Ballungsraumnahe Waldökosysteme", Berlin.

[12] Barsig, M. (2002). *Dissertation to Examinations of the Vitality of Oaks. Untersuchungen zur Vitalität
von Eichen (Quercus petraea und Q. robur) anhand von makroskopischen, mikroskopischen,
biochemischen und jahrringanalytischen Parametern.* Shaker Verlag, Aachen (Berichte aus der
Biologie) (Diss. TU Berlin). ISBN 3-8322-0375-3. Download as PDF-file (23 KB) (23 KB) (includes
only the English abstract) at

this research group received an award of the "Love Parade" for an MA Thesis about the damage of these acitivities in Tiergarten. The action group for preventing the construction of this tunnel in itself did not result in great influence, as the tunnel project went ahead as scheduled.

As for the Institute of Biology and Ecology at TU Berlin, the advisory practice regarding forests and trees was already established, and the request fitted into this existing practice. As noted, the mediator of the Science Shop and the researcher had previously co-authored a report in German and articles in the international literature, and the principle investigator continued to pursue an active research career in this field, which resulted in completion of his dissertation in 2002. In this case, the mediation by the Science Shop reinforced an existing relation between the university and the surrounding society by adding another dimension to it.

### 3.4.2  KREKO

The *kubus* researcher Wolfgang Endler (WE) reviously dealt with questions regarding cooperation and conflicts inside and between NGOs.  When the project Tiergarten Tunnel (case study 1) led to conflicts among the NGOs, the idea for the KREKO project was formed. In November 1996, WE initiated a "Zukunftswerkstatt" (future workshop, cf. Jungk & Müllert, 1989).  WE acquired two moderators (Michael Janßen (MJ) and his student MS) for this workshop. These three researchers together founded KREKO.

The project had several aims.  First, these researchers wished to address problems within and between NGOs, environmental groups and environmental associations, and also to improve internal communication and cooperation.  The project also wanted to be attractive to new activists, and revive the "we-feeling" among the activists.  It was also hoped to establish new forms of workshops and discussions for NGOs, environmental groups and environmental associations.   Finally, the project worked to promote pleasure in environmental-political engagement and to enforce the environmental-political influence. The main focus of KREKO was the competition and cooperation within and among environmental groups/NGOs and the support of constructive dealing with these problems, and also the formation of a network of these groups and relevant NGOs.

The project started in March 1997 with preparations and ended in March 1998 with the finalizing documentation.   The Stiftung Naturschutz Berlin (Berlin Foundation for Environmental Preservation) financed the project with € 3,000.  The funds were primarily spent on contracts for the services of three members of KREKO (workshop moderation and public relations) and on the documentation of the project (design and print).  In addition,

http://www.stadtentwicklung.berlin.de/umwelt/monitoring/en/download/barsig2002_abstract_en.pdf  or from http://www.stadtentwicklung.berlin.de/umwelt/monitoring/en/vegetation .

*kubus* and the BLN were involved with their own resources (i.e., the manpower of WE and MS).

The channels of communication were the regular project meetings.  Meetings for KREKO, dealing specifically with project negotiations, began in March 1997.  There was a clear role allocation in these meetings:  MJ and his student were the experts on workshop methods and organization, and WE, MS and another NGO member held the contacts to and the knowledge of the "NGO scene" in Berlin.  Cooperation in the group meetings, according to WE, MJ, and MS, was successful and constructive.

*Internet Visibility*
There was nothing to find at the homepage of the client group, at http://www.bln-berlin.de/. This was a project of a scholar with junior status at that time and a student. It focused on a social process and not on a cognitive outcome.  Within the institute the project had a low profile, although it belonged intellectually to the subject area of the research program.  The researcher wished to apply his expertise to enlighten the group of which he was a member. The pages related to this have been removed from the server of the TU Berlin, presumably because the researcher has left this institution in the meantime.  The report, however, was found to be mentioned previously in the university newspaper, at http://www.tu-berlin.de/presse/pi/1999/pi78.htm.

*Individual Researcher/Mediator Exposure*
The principle investigator, Michael Janszen, was referred to as the expert on workshop methods and organization.   He took part in the KREKO project as a private person, but was at the time also working for the project group "Cooperative project-development for communal health promotion," part of a research network related to the department of Public Health at the TU Berlin. The author is referenced with three articles in the bibliography of the case study report.  A search on AltaVista (+"Michael Janssen" +zusammenarbeit) revealed 37 hits, one of which included a newspaper message from the TU about this project at http://www.tu-berlin.de/presse/pi/1999/pi78.htm).

The director of the research group Professor Heiner Legewie is affiliated with the TU Berlin and a leading figure in this field.  PICA (in The Netherlands) lists six books with H. Legewie as author.  As found on AltaVista, Legewie maintains his own homepage at http://www.tu-berlin.de/fak8/ifg/psychologie/legewie/   as   well   as   at   http://www.medsoz.uni-freiburg.de/dkgw/whoiswho/personen/l/legewie.htm.   No  threads  to  the  project  were discovered at the second site, and the project could not be retrieved from the homepage of *kubus*.   Additionally on Legewie's homepage is a link to several articles, including one entitled "Kooperative Projektentwicklung im Gesundheits-, Sozial- und Umweltbereich."  This article is available as a Word-document download.

The third author of the book,[13] Birgit Boehm has a website at http://www.tu-berlin.de/fak8/ifg/psychologie/boehm/index.htm.

Neither Michael Janszen nor professor Heiner Legewie, can be retrieved in the *(Social) Science Citation Index*.  Several articles, however, are retrievable in German databases.[14] Birgit Boehm, the co-author of the book, is visible with an abstract in another domain of the psychology of labour and organizations.

*Conclusion*

As shown in the above visibility analysis, the principle investigator appears to function in the shadow of the Professor.  According to Professor Legewie, the Science Shop activities are intellectually interesting, but institutionally not important because they do not provide external funding.  The research of the institute is highly oriented to community-based research.  As for the participants, Michael Janssen left the university and Birgit Boehm received a promotion.  This accords with the conclusion of Michael Barsig in the first case-study that "the bitter experience was, that this kind of work does not get supported in a way one would like it. Nobody I know got a long-term job from this." As in the previous case, however, these researchers received a price for their work, that is, the Berliner Gesundheitspreis, 1998.[15]

The Science Shop can learn from this experience that one cannot rely on the homepage of individual scientists because the authors may not be on tenure-track and when they leave the university the visibility of the report can easily disappear.

### 3.4.3  WiLa Bonn

The main goal of the project was to create modules for a Germany-wide information, cooperation, and development network dealing with foundations in the field "Environment and Local Agenda 21."  Specifically, the project consisted of five sub-aims or modules.  One, to write and publish a compendium that gives an overview of the foundations in Germany that (financially) support groups and initiatives in the field of environmental protection and the Local Agenda 21.  The Local Agenda 21 has not yet been institutionalized in Germany, so there is still a need for funding.

---

[13] Böhm, B., Janßen, M., Legewie, H. (1999). *Zusammenarbeit professionell gestalten. Praxisleitfaden für Gesundheitsförderung, Sozialarbeit und Umweltschutz.* Freiburg: Lambertus-Verlag.

[14] Legewie, H. & Janßen, M. (1996): "Bürgerinitiativen fördern Gesundheit der Stadt." Veröffentlichungsreihe des Berliner Forschungsverbundes Public Health, 96-2; Janßen, M. und Legewie, H. (1998). Kooperation und Konflikt in der gesundheitsorientierten Stadtentwicklung. In: H. Heinemann. Stadtentwicklung und Gesundheit. Frankfurt/M.: Verlag für akademische Schriften, S. 83 – 112; Legewie, H. und Janßen, M. (1997). Bürgerinitiativen als Tätigkeitsfeld ehrenamtlicher Arbeit. In: Organisationsberatung. Supervision. Clinical Management 2, S. 151 – 163.

[15] Berliner Gesundheitspreis 1998, Kategorie III: Wissenschaft für Mensch und Gesellschaft. 22.04.1999 Ehrung für den Beitrag "Zusammenarbeit professionell gestalten: Ein Praxis-Leitfaden für kommunale Gesundheitsprojekte" (zus. mit M. Janßen und H. Legewie). Bundesweit ausgeschriebener Preis des AOK-Bundesverbandes und der Berliner Ärztekammer.

This project wanted to provide information on project funding by already existing foundations and moreover to encourage initiatives to build up one's own foundations.   A second objective was to establish a network of initiatives that are interested in building up a foundation, so these initiatives can support each other and be supported by the Science Shop.  This aim includes a platform on the Internet.  Thirdly, one aimed at conducting workshops on "founding a foundation," offered for initiatives looking into building up a foundation.  A fourth objective was to offer the conference "Foundations as Motors of the Local Agenda 21,"where varying sizes of foundations can interact.  A fifth objective was to write and publish a documentation of the conference as a textbook on "Foundations as project-agents for sustainability."

The scale of the project was defined by the duration of the funding. It ran from October 2000 until July 2002.   The project, as organized by the Science Shop, had several partners: Experts for the workshops on foundations, foundations that sent members to the conference and the workshops, the Bundesministerium für Umwelt, Naturschutz und Reaktorsicherheit—which funded the project—and persons, groups, or smaller NGOs who were interested in the topic as they planned to establish a foundation themselves.

The budget, which totalled € 120,000, was mainly financed by the Umweltbundesamt (UBA, Federal Environmental Agency), in arrangement with the Bundesministerium für Umwelt, Naturschutz und Reaktorsicherheit (BMU, Federal Environmental Ministry).  The Science Shop received about 10 percent of this funding.

*Internet Visibility*
As this project was extra-academic and semi-commercial in nature, an impact on the university publication system cannot be expected.  The Science Shop Bonn has a website at http://www.wilabonn.de/.  One of the headings on this site is entitled "Bücher zur Agenda 21," but this page does not lead to the publications from this project.  However, the reports can be found at http://www.wilabonn.de/stiftung.index.htm, at the link "Stiftungsnetzwerk."[16] These reports can be purchased or € 7.20 and 8, respectively.  The textbook and the compendium, outcomes from the project, were also found on this site.  The publication by Bücher and Valtentin is entitled "Stiftungen – Projektagenturen für Nachhaltigkeit." It can be purchased for € 8.

A follow up review volume, as mentioned in the case study report, can be found online at http://www.oekobase.de/html/beol.htm.   This can be considered as a very selective and honorific review of the project by an independent research team.[17] Thus, the project has an academic spin-off.

---

[16] Bühler, T. & Valentin, A. (Hrsg.) (2002): "Stiftungen - Projektagenturen für Nachhaltigkeit". Wissenschaftsladen Bonn; Bühler, T. mit Beiträgen von Valentin, A. & Janenz, S. (2001): "Projektförderung durch Stiftungen - Umweltschutz und lokale Agenda 21". Wissenschaftsladen Bonn.

A search for "Fundraising Academy" resulted in 10 hits. The website of this academy is located at http://www.fundraising-akademie.de/, and the start of the academy is noted at http://www.ekd.de/bulletin/bulletin_2791.html. The Fundraising Akademie can also be found at http://www.sozialmarketing.de/akademie.htm. Additionally, an overview of Spendenrat, etc., can be found at http://www.fundraisingbuero.de/a/text/links.htm. The website of the Deutscher Fundraising Verband e.V./German Fundraising Society is located at http://www.sozialmarketing.de/bsm_engl.htm. This organization is extremely well-networked and visible on the Internet. However, the collaboration with the Science Shop is not mentioned on any of these websites.

*Individual Researcher/Mediator Exposure*

Two publications by Theo Bücher, the researcher at the Science Shop who was responsible for this project (chairman of the association, managing coordination) are listed in the references. Theo Bücher is most likely the interviewee for the Science Shop Bonn.

In relation to the report, it is difficult for an outsider to understand who the client is. The interview partner, WS, is one of the voluntary members of the NGO, who took part in the workshops as well as the conference. Any further information is difficult to find. Additionally, the lecturer has wished to remain anonymous in the interviews. Regarding the client organization, the case study report claimed that the NGO is well-documented on the Internet. The website mentioned in the report at
http://www.agenda_buero_koepenick.bei.t-online.de does no longer exist. The project, however, is well-documented at
http://www.berlin.de/ba-treptow-koepenick/Aktuelles/1AInhaltsverzeichnis.html.
The brochure of 210 pages is also available in hardcopy.

In searching the Science Citation Index, T. Bücher and S. Janenz had no publications. A. Valentin had four listed, but not in this area. In summary, no international publications were found, and a PICA search returned no results. As noted, this is not amazing given the professional aspirations of these authors. The WiLa Bonn wishes primarily to offer expertise and to be of service on a market. Given the extra-university location, publication is not explicitly rewarded. As noted, the publications, however, were mentioned in an academic review that attempts to indicate high-quality contributions to the research area.

---

[17] "Die Arbeitsstelle für Ökologie und Pädagogik an der Freien Universität Berlin leistet hier Hilfestellung. Von der Arbeitsstelle wird das Angebot des Marktes gesichtet und nach einheitlichen, überschaubaren Kriterien bewertet. So wird aus der Fülle der Neuerscheinungen eine Auswahl jener Materialien getroffen, deren Qualität nach unserer Auffassung über dem Durchschnitt liegt. Damit ist "Ökologie und Lernen" kein Rezensionsorgan im herkömmlichen Sinne. Denn hier wird nicht jede "Medieneinheit" besprochen, die wir in diesem Feld registriert haben. Vielmehr handelt es sich um eine gezielte Selektion - eben um Empfehlungen."

*Conclusion*

The project had primarily a function in the professionalization strategy of the WiLa Bonn Science Shop. The results provided valuable information and contact sources in relation to the project development and funding, as well as improvement in the contacts with foundations and other funding institutions (such as the Bundesministerium für Umwelt, Naturschutz und Reaktorsicherheit). In addition, these contacts are being used to facilitate the work of the Science Shop. However, a feedback on the research and/or higher-education system could not be found.

### 3.4.4  Conclusions and Recommendations

The German Science Shops under study operate professionally. There is a strong relation with the alternative scene in Germany. However, the visibility of these activities is not high because the clients have no great interest in using the reports other than as instruments in their action perspectives. Thus, the attractiveness to larger audiences—which may struggle with similar problems—is not further exploited. The university (in this case, the TU Berlin) could use the Science Shop, the reports, and the mediation more systematically to make some of the strengths of its research programs visible in the social contexts. The Science Shops make an effort using their web pages, but a further integration of these activities in the PR of the university might help to improve these efforts.

Perhaps, a further integration of the Science Shop activities into the higher education system should also be considered. Why are so few students involved? In other words, we are able to see a highly motivated effort using academic standards, but the efforts are perhaps not sufficiently made clear to more general audiences. In all three cases, important work has been done that is recognized in the relevant scientific community, but not in the wider context where this work may also be relevant.

## 3.5    Spanish Case Studies

Three different Science Shops were a part of the case studies in Spain.  The first, Pax Mediterranea, was founded in 1995 and is located in Seville, Spain.  It deals with research within the social arena of ecology, economical development and social cohesion strategy with environmental and socially sustainable perspectives.  It is actively involved in various European and local research and monitoring projects and observatories.

The second is ACS.  It was founded in 1994 and is affiliated with two major architectural schools in Seville.  The members' main concerns are social instruction in the universities, the construction of a sustainable habitat in inner cities, equality on a global scale and the instruction of citizens, rather than only trained architects.

Finally, Science Shop ISTAS is a self-funded technical foundations promoted by the Spanish Trade Unions Confederation (CC.OO). It supports social activities for the improvement of working conditions and environmental protection in Spain.  It was founded in order to back trade union action in the field of occupational health and environmental protection.

### 3.5.1  Urban Ecology Strategy Design, Seville 2025

This project resulted from the desire of a group of "green" organizations wanting to do an independent study on the ecological issues that were present in Seville society. These groups organized in order to look at possible future scenarios where their input and action could be called for.  Furthermore, it was important to create an Agenda 21 for Seville from the date of the study (1999) to the year 2025.  The group contacted the Science Shop Pax Mediterranea, though Teresa Rojo, a member of Ateneo Verde (one of the NGOs included in the grouping), and also a staff-member of the Department of Sociology at the University of Sevilla.

The NGOs involved include Ateneo Verde and the Los Verdes Party.  Ateneo Verde is a cultural project site which was opened in order to defend ecology, peace, solidarity, and human rights.  Los Verdes de Andalucía is a political party that is a part of the global Green Party movement.  The research consisted of supervised open-brain-storming workshops in order to create a debate within the organization as well as with other interested parties.  The project lasted for three months, and the costs were met by the NGO.

*Internet Visibility of the Project*
The website at http://talike.fie.us.es/ateneoverde/ did provide information about Ateneo Verde, but no longer exists.  Page http://verdes.es/andalucia/quien.php provides information about the Green political party and its activities.

At http://www.loka.org/techniques.htm[18] a publication can be retrieved by Teresa Rojo and Alain Labatut, the researcher and the Science Shop mediator of this project, respectively.

*Individual Researcher/Mediator Exposure*

Teresa Rojo did the work for this project as an independent volunteer, as her university does not support this type of service.   The homepage of Teresa Rojo is located at: http://investigacion.us.es/sisius/sis_showpub.php?idpers=2876.   This site lists a book plus the conference paper on the project.[19]   Teresa Rojo is also listed as one of the national monitors for Spain on the site of the European Awareness Scenario Workshops at http://www.cordis.lu/easw/src/monitor2.htm#E.  European Awareness Scenario Workshops were used as the methodology in the project under study.[20]

Juan Maestre holds a chair position and is Professor in the Department of Sociology, along with Teresa Rojo.  He is responsible for the link to the publication of the end results of this project.  His home-page can be found at http://investigacion.us.es/sisius/sis_showpub.php?idpers=2909.    He has no publications listed on this site, only work that he has supervised.

The home page of the NGO representative, Ricardo Marques Sillero is at http://otri.us.es/sisius/sis_showpub.php?idpers=3220.   Sillero is a Professor in electronics and electromagnetism.  He is the leading scientist in his field, with 55 papers listed in the *Science Citation Index*.    He  can  also  be  found  at  http://www.izquierda-unida.es/Entes/comu12.htm.  An action group for the elaboration of a new social contract between the university and society, subscribed to by both Rojo and Marques, can be found at http://www.us.es/compromiso/somos.htm.

As noted, a search for Sillero returned 55 papers in the *Science Citation Index*.  A search for publications  by  Rojo,  however,  returned  no  results.    A  PICA  search  was  equally unsuccessful, and a search on amazon.com and also on the French alapage.com, no results were found.

*Conclusion*

The project seems the result of a densely connected network among Sevillan intellectuals. The results are interesting, published, and well disseminated. Both the theoretical book and the methodological results are impressive. Through local relations the reception seems also to be secured.

---

[18] The Loka Institute can be considered as a U.S. Science Shop (Sclove, 1995).
[19] Rojo, T.: *Sevilla 2010, Metrópoli Ecológica. Aplicación de la Metodología EASW*. 202 Páginas. Universidad de Sevilla. 2001. ISBN/D.L.: 84-600-9628-9; Rojo, T.: "La Sociología Prospectiva incorpora instrumentos participativos". *VII Congreso Español de Sociología*. Federación Española de Sociología. Salamanca. 2001.

### 3.5.2  Architectural Study for Romany Community, "Los Perdigones"

The project began and was carried out in 2000, when the NGO Human Rights of Andalusia were informed that a Romany (Gypsy) shanty neighbourhood was being removed from public/private land in order to make room for building contractors.  This NGO contacted the Science Shop Architecture and Social Commitment.  The scientist/university involved was Durán Rojo Architects.  The methodology consisted of the organization of a contest in order to provide alternative solutions for housing adapted to Romany social requirements, and, in finding this solution, the avoidance of expulsion of the community.  The project had a duration of one and one-half months, and involved a number of students and professional architects.  The minimal costs of the organization were met by the NGO and the Science Shop.

The NGO's founding principles mirror the Universal Declaration of Human Rights as declared by the United Nations in 1948.  It is mainly affiliated with the Andalusia territory, but it also is involved in global concerns.  Apart from educational activities and active participation in human rights concerns, the NGO provides support to victims of human rights violations in Andalusia, particularly to those who have no resources.

*Internet Visibility of the Project*
The website for the Science Shop is located at http://www.arquisocial.org/ACS.htm, and the client organization can be found at http://www.apdha.org.  A link with theoretical background through John O'Connell, University College, Cork, Ireland, can be found at http://www.ucc.ie/ucc/units/equality/pubs/Minority/oconnell.htm.
http://www.fidas.org/boletin/fidas26.html provides information as well as a picture about the project.

On the basis of the follow-up activities mentioned in the case study report, a site was found for Ventura Galera, an architect who mediated the project at
 http://www.sevillaqueremos.org/documenta/ventura.htm.  This page provides a description of the city planning.  A special issue of *FIDAS*, the journal of the *Fundación para la Investigación y Difusión de la Arquitectura*, about this project is made visible at http://www.fidas.org/boletin/numero26/0indice26.html. The Table of Contents of the special issue is provided.

A relevant website of the dependencias municipals de Carmona can be found at
 http://www.carmona.org/dependen.htm.

---

[20] European Awareness Scenario Workshops (EASW) were also used as a methodology for the national dissemination workshops in INTERACTS, WorkPackage 5.

*Individual Researcher/Mediator Exposure*

The scientist involved, Jesus Rojo Carrero, is not mentioned further. The winners of the contest were students at the time, but have founded now their own company. However, they are not visible on the Internet.

*Conclusion*

Twelve workable designs were submitted for the Romany Community housing project. None of the interviewed individuals are presently using the results in their curriculum or professional lives. The special issue of *FIDAS* can be considered as a wide dissemination in professional circles. The project is thereby archived and visible on the Internet.

### 3.5.3  Health and environmental hazards at cement kilns waste incineration

In early 2001, the issue of burning cattle meat in kilns became a risk concern for both workers and for the environment. ISTAS made a recommendation to CCOO to study the issue, in the view of minimizing worker environmental risk. The research project addressed this issue as discovering such risks. The methodology involved data gathering and analysis, through investigating research from incineration studies, particularly those carried out in kilns and of incinerating animal meat. The duration of the project was two years and the costs included human resources including one full-time technician for the duration of one year.

The NGO is a union organization and umbrella body for State Federations, National Confederations and Regional Unions. Its main objective is to counteract all forms of discrimination and capitalist oppression. Carlos Martinez was interviewed as the representative of the CCOO, and Juan Romero Agud represented an anonymous NGO at the grassroots level.

*Internet Visibility of the Project*

Information on ISTAS (Instituto Sindical de Trabajo, Ambiente y Salud) can be found at http://www.istas.net. ISTAS is the research institute of the CCOO that provides these Science Shop mediations.

The CCOO can be accessed at http://www.ccoo.es/pdfs/estatutos.pdf. The Union Guide (*Guía Sindical para el seguimiento y prevención de riesgos derivados de la coincineración en cementeras*--Union Guide for the follow up and prevention of risks derived from the incineration at cement kilns), published by CC.OO and CC.OO-Fecoma, 2002, is also located here.

Scientists for the Elimination of Toxic Pollutants have a website at: http://www.istas.net/decops.htm. In this context the *Stockholm Agreement on Persistent Organic Polluting Agents: An international instrument for a global problem* was written by Estefanía Blount who is the mediator of the Science Shop in this project. This document can be retrieved at http://irptc.unep.ch/pops/.

At http://www.anped.org/PDF/11spaccleanprsclist.pdf one finds information about clean production technologies.  However, neither the United Nations Environmental Program websites at *.unep.ch nor the links to the Stockholm Convention on Persistent Organic Pollutants did function at the date of this research (February 2003).

The Guia Sindical which resulted from this project, is described on p. 15 of the document at http://www.istas.net/ma/daphnia/daphnia28.pdf.  This Guide itself is directly available on the Internet at http://www.istas.net/portada/guiacemen1.pdf.

*Individual Researcher/Mediator Exposure*
Information on the scientist involved can be found at http://www.ivia.es.  Fernando Pomares Garcia, the team leader for fertilization and soil conservation, does not have a personal page, but is a member of the staff at the Instituto Valenciano de Investigaciones Agrarias. The researcher works in the department of natural resources of this institute.

The researchers and the Science Shop mediator (Estafanía Blount) are both on the participant list of the third national conference of environmental science in 1996, located at http://www.conama.es/iiiconama/.  The report of Miguel Crespo can be found at
 http://www.ecologistasenaccion.org/accion/residuos/ponencias/4-1-Miquel.doc.

Fernando Pomares has 14 papers in the *Science Citation Index* since 1988.  This is an average publication profile for an academic, but this output can be considered as high for an author in the setting of a public research institute. (These institutes are largely contract-driven.)

*Conclusion*
This project is a beautiful example of an extra-university activity that illustrates the "knowledge-based society." The diffusion of results shows the success of this case study since the client has continued the process by editing and distributing most of the outcomes. It has allowed the mediating agency of the Science Shop to reach a social group in need of scientific assessment.

### 3.5.4  Conclusions and Recommendations

- Bring these rather heterogeneous activities together by actively using the Internet (hyperlinks).
- Show the mechanism of organizing social capital.
- Exhibit the in themselves impressive results in a more coherent framework.

## 3.6 UK Case Studies

The Liverpool case studies are part of the long-term investment of two senior researchers (partners 6 and 7 of the INTERACTS consortium) in community-based teaching, learning, and research. The project can therefore be considered as methodologically driven by the research program. The two researchers, Irene Hall at Liverpool Hope University and David Hall at Liverpool University co-authored the book *Practical Social Research: project work in the community* (Basingstoke, etc.: Macmillan) in 1996, which details their methodology. This book is available with six copies in the inter-library loan system of the Netherlands, of which three copies can be located within university libraries.

David Hall is visible both in the social science literature and on the Internet, while Irene Hall is less visible in this literature. Her strong focus on community-based research leads to a engaged involvement in publications that extend beyond the academic community. David Hall, however, is a more prolific author in the social science literature. Using the library of Hall's home university, two of his previous books from the 1970s could be retrieved,[21] and two publications are included in the *Social Science Citation Index* in 1991.[22] The author maintains a website based on his membership in the Globalization and Social Exclusion Unit, Department of Sociology, University of Liverpool.

A new book by these two authors entitled *Evaluation and Social Research* is forthcoming with Macmillan in 2003. The collaboration that carries the Science Shop projects in Liverpool has also been embedded and disseminated in the CoBaLT project (Community-Based Learning Teamwork). The CoBaLT project is a three-year project funded by the Higher Education Funding Council for England under its Fund for the Development of Teaching and Learning. CoBaLT is a Consortium project involving Sociology at Liverpool Hope, the Department of Sociology, Social Policy and Social Work Studies, University of Liverpool, and the Department of Cultural Studies and Sociology at the University of Birmingham.

In summary, this program of community-based research is methodologically driven by a theoretical approach that focuses on learning from interactions and the researchers involved are deeply committed to the objectives of community-based learning. In an existing framework of collaboration with volunteer organizations and charities, evaluations of arrangements in the health sector are organized into student projects that function in a specific disciplinary framework. The Science Shop model suits this framework because the mediation model potentially adds structure and legitimation to the ongoing activities. The name "Science Shop", however, is not essential to the enterprise.

---

[21] Hall, David J. and Margaret Stacey (Eds.) (1979). *Beyond separation: further studies of children in hospital.* London: Routledge & Kegan Paul; Hall, David J. (1977). *Social relations and innovation: changing the state of play in hospitals.* London: Routledge & Kegan Paul.

[22] Hall, D. (1991). Book review of "Choosing For Children - Parents Consent To Surgery" by P. Alderson, *Sociology Of Health & Illness*, 13 (3), 433-434; Hall, D. (1991). The Research Imperative and Bureaucratic Control - The Case Of Clinical Research, *Social Science & Medicine* 32 (3), 333-342.

The UK case studies involved two of such "Science Shops:" *Interchange* and *Student Link*. Interchange was established as a registered charity in 1994, as the result of a merger between two organisations, Merseyside Community Research Exchange (MCRE; initially funded through the Enterprise in Higher Education initiative of the UK Department for Education and Employment) and the Liverpool Science Shop (established with a grant from the Nuffield Foundation). MCRE, which began work in 1991, was inspired by the earlier model of the Manchester Research Exchange (www.commex.man.uk/commex/home.htm ), and involved students from all three local universities - Liverpool University, Liverpool Hope University College and Liverpool John Moores University - undertaking research projects with local NGOs. The Liverpool Science Shop had provided access to scientific information at Liverpool University through one-off requests and student projects, and the merger was intended to provide a more coherent service to the community.

Student Link at the University of Wolverhampton is the comparison Science Shop example. It enables final year undergraduate students to conduct applied research projects for one semester (15 credits) or two semesters (30 credits).  The objectives of Student Link are: to provide organisations with a forum to gain access to additional skills which will support them in their work; to give students the opportunity to develop research, vocational and personal and transferable skills in a practical and useful way; to match student undertaking independent study with organisations wanting project work carried out; to enable students to be assessed as part of their academic programme on a range of skills demonstrated outside the academic context; and to enable students to evaluate their own learning and skills development in the context of an organisationally based project.

### 3.6.1  Benington Hospital

Four separate projects were conducted under this case study.

1   The Volunteer Scheme: its Role in Nurse Recruitment.  Participants included three final year graduates, a supervisor from Liverpool Hope University College, and a volunteer manager.  The main aim of Project One was to assess the benefits of volunteering for student nurses.
2   "Befriending and Counselling in Accident and Emergency (A+E)," involved a Master's student, a Supervisor from Liverpool Hope University College, and a Family Support Manager. The main focus of this project was to evaluate the Befriending service (volunteers who support the relatives of patients), as well as an investigation of staff stress, and of the Trauma and Bereavement counselling service.
3   Volunteers and Infection Control consisted of one final year undergraduate, a supervisor from Liverpool Hope University College, a Volunteer Manager, and a Nurse Consultant.  The aims were to evaluate infection control practices by hospital volunteers and to identify volunteers' fears/ concerns about infection control and to identify their information needs.

4   Senior nursing staff and infection control, utilized a Masters student, a Supervisor from the University of Liverpool , and a Nurse Consultant.  The aims were to explore the role of opinion leaders and decision-makers in providing suitable care in relation to infection control.

All four projects relied on semi-structured interviews, using schedules with open-ended questions to supply the bulk of the data. Self-completion questionnaires were also used, and all students kept diaries of their observations and experiences. Observation provided orientation to the research, and featured more in the students' reflective accounts rather than in the client reports. The undergraduate projects were conducted between October and May (8 months), the postgraduate projects between December and September, 2000 (10 months).  The Volunteer Scheme registered all students as volunteers, and covered travel expenses, photocopying and the ordering of journals.

*Internet Visibility of the Project*
The Times Higher Education Supplement article mentioned in the INTERACTS case study report can be retrieved at http://www.thes.co.uk/subscriptions/free_trial/main.asp.   This article features an interview with a student and the supervisor.  Other referrals are made to *Evaluation and Social Research,* Palgrave, 2003, and also a reference is made to the report *Hospital Volunteers' Perceptions and Understanding of Infection Control: An Exploratory Study* (2002).  Neither one of these publications were retrievable on the Internet.

The case study report mentions also two articles appearing in the *British Journal of Infection Control.* This is the scientific journal of the Infection Control Nursing Association of the UK. The journal can be accessed from the home page of the association at http://www.icna.co.uk, but the article mentioned could not be found.  An article by Arnold and Rice pointed out could also not be found.  AltaVista was searched for "+arnold +rice" and well as "Liverpool Hope University College," but without results.

*Individual Researcher/Mediator Exposure*
The authors of all three case study reports are Irene and David Hall.  As noted, David Hall's home page can be found at http://www.gseu.org.uk/people/hall.htm. This is part of the website of GSEU: Globalization and Social Exclusion Unit, for the Department of Sociology at the University of Liverpool.  However, at this website no publications related to the projects can be retrieved.

A search on AltaVista using "Practical Social Research" (a Hall publication) returned 29 hits. One link, http://www.hope.ac.uk/cobalt/CBL%20at%20Liverpool.html, mentions the (Hall and Hall, 1996) book as a textbook under the title "This is Practical Social Research." This is part of the Cobalt project mentioned in the report as a source.  The book is additionally mentioned at the Macmillan site, at
http://www.palgrave.com/catalogue/catalogue.asp?Title_Id=0333606736.   The University of Liverpool Library provides the book as a hit, at

http://library.liv.ac.uk/search/a?SEARCH=Hall%2C+Irene.   This library also includes two books by David Hall, one published in 1977 and one is 1979.

An additional search was performed on http://www.palgrave-usa.com/ for the forthcoming book, *Evaluation and Social Research*, but nothing was yet retrieved.  David Hall has one article and one book review in the *Social Science Citation Index* (in 1991).  Irene Hall's work could not be retrieved using this database.

### 3.6.2  Lakeview Day Centre

The main aim of the project was to provide an independent evaluation of a day centre for older people, from the service-users' perspective.  The project began with a request from the NGO for an external evaluation of the Day Centre.  The request eventually found its way to the office of Interchange, and from there, the project was listed by Interchange as a possible student research project. Two undergraduate students chose this subject for their applied social research project.

The process of negotiation began with the visit of the Interchange Co-ordinator to Lakeview Hospital to discuss the proposed project with the Chief Executive and the Day Centre Manager. This was followed by the students visiting the Hospital and the Day Centre to negotiate and plan the project. When these initial negotiations had moved to a shared view of the project, the academic supervisor also attended a meeting at the Hospital with the students and the Chief Executive to reach final agreement on the nature and scope of the research. This is summarised in the negotiation agreement signed by all parties.

Data collection was largely accomplished by questionnaire and interview.  The student researchers visited the Day Centre on 13 separate occasions between October and January, spread across the different days of the week, when there were different service users and different volunteers on duty, thus ensuring coverage of the range of service users and service provision.

Interviewees for the case study included two final-year undergraduate students, the Chief Executive of Lakeview Hospital Trust, the student supervisor in the Department of Sociology at the University of Liverpool, the coordinator of Interchange, the Science Shop, and the Dean of Faculty and the Senior Academic Manager at the University of Liverpool.  The project began in October 1999 and was completed in May 2000.  Lakeview Hospital paid for the students' travelling expenses, and for multiple copies of the final report.

Lakeview Hospital Trust is a non-profit making charitable organisation providing social and medical care to older people. The Hospital itself has been in existence since the early 1900s, first as a voluntary institution, then as part of the National Health Service from 1948, but it has been an independent charitable trust since 1983.  A number of services are provided on site, all for the care of the older person, that is, people who are of pension age and older.

*Internet Visibility of the Project*

The project could not be made visible on the Internet. An article by Fong and Cronin, the two final-year undergraduate students, was mentioned in the report.  Searches in AltaVista including +fong +cronin, +fong +cronin +liverpool, +fong +cronin +liverpool +sociology, and +fong +cronin "a cottage industry of care" retrieved no hits.  Lakeview Hospital mentioned on page 20 was searched for using "+lakeview +hospital," "+lakeview +hospital +england," "+lakeview +hospital +liverpool," and "+lakeview +"day centre"."  Nothing about this project could be located.[23]

*Conclusion*

The findings, summarised from the report's conclusions, are as follows:  One, the club provides a valuable service, keeping many of the people that go from slipping into complete isolation, and provides a caring, supportive and fun day out.  Two, the volunteers create a relaxed and friendly atmosphere each day of the week, and the club provides a community feeling and network of friends that would otherwise not exist.  Three, the services provided by the club are very much appreciated and enjoyed by the service users. Every effort is made by the volunteers to accommodate individual needs and preference. Any criticisms found are minor and based on individual tastes rather than any specific problems with the club.  Four, the transport service is invaluable, efficiently run, and provides a safe means of getting to and from the club.

This project provides us perhaps with the extreme case of the objective of social capital formation and practical experience for students as the main objectives. The scientific objectives are less clearly specified. The outcome is professionally oriented.

### 3.6.3  Midlands Befriending Service

The project was part of an ongoing relationship with Age Concern and originated in a pilot scheme set up in 1996, as a development of Student Link. The Student Link Age Concern Scheme pilot was reported as differing from Student Link generally because several students could be allocated to the organisation to work in a team, and there was to be a rolling programme of research rather than the normal one-off research project.

The participants in the project consisted of one final-year undergraduate student at the University of Wolverhampton, the academic supervisor, the coordinator of Student Link, the Science Shop, the Befriending Service coordinator, and the manager of the NGO.  The project began in January 2002 and was completed in May 2002.  The NGO registered the student as one of their volunteers in order to cover insurance needs, and provided for travel expenses and postage up to a previously agreed-upon budget.  The project had the specific aim of producing a review and evaluation of the Befriending Service for the NGO.  This was accomplished by discovering the views of the clients who had received the service in their homes, as well as the views of the volunteer visitors.

---

[23] Only after finishing this report, we were informed that the names of the clients were pseudonyms in order to guarantee their anonymity.

The pilot scheme involved considerably more initial liaison between university tutors and Age Concern staff to determine the area of provision to be evaluated than would normally be involved in a one-off project. The benefit was that in succeeding years, tutors would be able to spend less time in negotiation. The research involved the student visiting the premises of the NGO on a weekly basis for around 6 hours per week, where she could meet with volunteers and attend their training sessions, devise her questions for the interviews and conduct interviews with clients and volunteers.

Age Concern Midlands is an independent charity and limited company which targets local people over 50, who are frail or who are suffering from loneliness. The Befriending Service is one of a number of services and provides a volunteer visiting service to older people who have recently returned home after being discharged from hospital or from a Social Services (local authority) residential Community Resource Centre.

*Internet Visibility of the Project*
A search for the Midlands Befriending Service (AltaVista, +midlands befriending service +age concern), provided a link to
http://www.wolverhampton.gov.uk/socserv/oldir/advice_information1.htm, which
gives information about Wolverhampton services.  Also http://www.ageconcern.org.uk/ is a website that provides a description and information about the age concern charity.  A specific description can be found at http://www.ageconcern.org.uk/ageconcern/about.htm.

*Individual Researcher/Mediator Exposure*
On amazon.co.uk, a book was listed with a contribution by Hall and Hall, entitled "Academic and Educational Development: Research, Evaluation and Changing Practice in Higher Education (Staff and Educational Development)."  Additionally, a study by Irene Hall was retrieved, entitled "Community action versus pollution: a study of a residents' group in a Welsh urban area."

*Conclusion*
The report has been used as a guide to incoming coordinators. This report could not be retrieved. It is not clear from the case study reports whether the student reports are publicly available.

### 3.6.4  Conclusions and recommendations

In our opinion, these researchers and Science Shops in the UK may not have paid sufficient attention to the function of the projects in a wider context, for example, by making the results available at the Internet (if needed after anonymization). The project results are disseminated locally, while they are guided from a theoretical program. We assume that the researchers themselves use materials from the projects in their scientific publications and books, but the students seem not to be stimulated to bring the reports to the attention of a wider audience. The practical emphasis may cause legitimation problems in the theoretical context of university research and higher education.

Perhaps, one should evaluate the efforts primarily as an outcome of higher-education and community servicing without necessarily the ambition to achieve scientific status and visibility from this work. However, it seems to us that the projects may be more valuable than this as exercises in generating social practice. In addition to publishing the student reports on the Internet, the researchers can perhaps report more frequently about the results in the literature.

## 3.7 Romanian Case Studies

There are two Science Shops involved in the Romania case studies. InterMEDIU Information, Consultancy and ODL Department, at the Technical University of Iasi, was responsible for case studies one and two. InterMEDIU was founded in April 1999 as a non-profit, independent department of the Technical University of Iasi. It is based in the Faculty of Industrial Chemisty, as a result of the bilateral cooperation agreement with the University of Groningen in The Netherlands. This cooperation is also known as the MATRA program.

InterMEDIU is self-financed through its projects. The InterMEDIU Centre cooperates with organizations in the civil society in relation to matters involving information, consultancy and research in the field of environmental protection, as well as education and training.

Specifically, InterMEDIU is involved in the fields of environmental protection, environmental management, and environmental education and awareness programs. The Science Shop is seen as an interface between the university and society, as its main objectives are related to the transfer of knowledge in the field of environmental protection from the University toward civil society structures, the facilitation of public access to environmental issues and contribution to capacity building of environmental groups. Additionally, the Science Shop offers the organization of student programs and continuous education programs related to environmental education as well as student experience. Students find their cooperation with the Science Shop very useful, as they are often interested in translating a real-life problem into a scientific research proposal.

The second Science Shop is InterMEDIU Information and Research Centre, "Al.I.Cuza," at the Faculty of Biology, University of Iasi. This Science Shop intermediated in case study three. It began in March 1999 as a result of the above mentioned MATRA program. Two main motivations supported its formation: one, it hoped to place biological scientific knowledge at free disposal for non-profit organizations and other groups that lacked funds for scientific research. Two, it was aimed at establishing a tighter connection between academic education and research and societal needs. Since the Science Shop is mainly focused in the area of biology, its present and future tasks center around nature protection. This involves two aspects: the protection of threatened populations or species, and the protection of their habitat.

Resulting research involves areas such as water, air, and soil pollution, drinking water supply, waste management, population health aspects, energy efficiency, landscape, and biodiversity. As part of the Science Shop activity, students are taught how to contact people and institutions and formulate scientific information in a foreign language. The Science Shop additionally provides environmental awareness programs for younger students and also contributed to the established of the European Centre of Excellence for European Studies, with a focus on regional development at "Al.I. Cuza" University in 2000.

### 3.7.1 Evaluation of the quality of drinking water supplied in the city of Iasi

This study (1999) represents the pilot project of the Science Shop InterMEDIU (Technical University of Iasi). The project was initiated through discussions between the Science Shop, representatives of the NGO, the Dutch partners of the MATRA program, and the staff of the Environmental Engineering Department. The NGO involved in this project was the Academic Organization of Environmental Engineering and Sustainable Development. This is a Romanian legal entity. The aim of the NGO is to promote ecological behavior and technical solutions, education and awareness of the public in order to understand the need of sustainable practices and the conservation of the environment.

The aims of the project included the following: to consult the community about the drinking water, to compare the qualitative problems raised about the situation in the treatment plants, to formulate proposals for improving the existent situation, and to organize a public debate concerning the drinking water quality. An assessment of problems related to drinking water was analyzed through the use of 2584 questionnaires, addressed to the populations living in the neighbourhoods in question. The results of these questionnaires served as a point of discussion of treatment technologies currently applied by the Water Works company for different sources of raw water.

*Internet Visibility of the Project*
Descriptions of the Science Shops in Romania can be found at
http://www.fwn.rug.nl/chemshop/roemeng.html and at
http://www.ssc.unimaas.nl/LSW/news_rom.htm. The "ODL department" of the Technical University of Iasi is visible at:
http://www.tuiasi.ro/pages/TUI97/ch97.htm#ENVIRONMENTAL_DEP
Carmen Teodosiu is listed among the staff of the department, but InterMEDIU does not seem to have its own webpage. While Teodosiu is retrievable, MATRA or Henk Mulder (the coordinator of this program) are not visible in this context. The collaboration is somewhat visible on the Internet (in less than 10 hits) through the Dutch Science Shop pages, but not on the Romanian side.

The NGO is the publishing house of the recently-created *Environmental Engineering and Management Journal*. One of the publications of this project, focusing on the role of the Science Shops, is found in this journal. It is entitled "Science Shop contributions to environmental curriculum development," and it was written by C. Teodosiu and A. F. Caliman.[24] The journal itself, which transpires a high level of ambition, can be retrieved with nine hits. The homepage is located at http://omicron.ch.tuiasi.ro/EEMJ/ . The website http://omicron.ch.tuiasi.ro/iceem/ leads to part of a special issue reporting from the first

---

[24] Teodosiu C., Caliman A.F. (2002). "Science Shop contributions to environmental curriculum development", *Environmental Engineering and Management Journal*, 1(2), 271-293.

International Conference in Environment Engineering and Management, 26-28 September 2002.

The journal *European Water Management* is not covered by the ISI-databases, but the article by Teodosiu and Stanciu-Petrea ("Evaluation of the drinking water quality and quantity in the City of Iasi, Romania") is retrievable in the Dutch Inter-library system PICA.  The article is available at five universities.   The homepage of the journal is: http://www.ewpca.de/journal/journal.htm.  Further, the article was announced in a European Water Management News (press release) at
http://www.riza.nl/ewa_news/news_22_august_2001.html.

An additional publication authored by Caliman, Teodosiu, and Balasanian was retrieved, entitled "Process conditions for industrial wastewater treatment by heterogeneous photocatalysis."

*Individual Researcher/Mediator Exposure*
The project involved the NGO, Academic Organization for Environmental Engineering and Sustainable Development.  The supervisor of the project (Carmen Teodosiu), a senior staff member of the Department of Environmental Engineering and at the same time the science shop manager, and a science shop research (Monica Stanciu Petrea) were directly involved in the project realization. The Chemistry Science Shop coordinator at the University of Groningen was also involved, as well as ten students of the Faculty of Industrial Chemistry, with a specialization in Environmental Engineering.  The students were in their third or fourth year of study.

One of the students (M. Hristea) is coauthor of an article with Teodosiu, "Possibilities to upgrade the conventional drinking water treatment technology," found in *Buletinul Institutului Politehnic Iaşi*.  A report (in collaboration with the Dutch partners) is entitled "EMS in factories. The introduction of Environmental Management Systems in the metalprocessing and ceramic industry in Romania and The Netherlands (Engelstalig)" and can be retrieved at
 http://www.fwn.rug.nl/chemshop/cwpub.html.

Two articles were retrieved in the *Science Citation Index* using the name of C. Teodosiu.  One is authored by Teodosiu, Kennedy, Van Straten, and Schippers.  It is entitled "Evaluation of secondary refinery effluent treatment using ultrafiltration membranes," and located in the journal of *Water Research*.[25]  The other one is C. Teodosiu, O. Pastravanu, and M. Macoveanu (2000), Neural network models for ultrafiltration and backwashing, *Water Research* 34(18), 4371-4380.[26] The other authors of the report were not retrievable in the ISI database.

---

[25] Teodosiu CC; Kennedy MD; Van Straten HA; Schippers JC (1999). Evaluation of secondary refinery effluent treatment using ultrafiltration membranes, *Water Research*, 33(9), 2172-2180.

[26] Teodosiu C; Pastravanu O; Macoveanu M (2000). Neural network models for ultrafiltration and backwashing, *Water Research* 34(18), 4371-4380.

*Conclusion*

A correlation between the technical conditions and the degree of treatment was realized as well as recommendations for improving the existing situation were provided. The project received good media coverage and the students of the Environmental Engineering Department were provided with an opportunity to apply their knowledge relating to Water Treatment technologies, and also to learn more about the techniques of social inquiry, project management, and computer applications.

### 3.7.2 The impact of wastewaters from the industrial production of yeast on the river of Siret

The project began as a request from an environmental NGO, the Ecology and Tourism Club Moldavia. Founded in March 1995, its mission relates to the promotion of principles regarding the real and active protection of the environment. The project's objective was the evaluation of the environmental impact of the wastewaters generated from yeast production on the receiving waters of the river Siret. The major objectives included the evaluation of the industrial process of yeast fabrication from molasses, with respect to emissions in wastewaters, their discharge and treatment possibilities; analysis of the environmental impacts produced by wastewaters considering their possible dischange into the sewage system without preliminary treatment; and suggestions for improving the existent situation. The project's duration was three months (February, March, and May 2000), and all costs were covered through the MATRA project funds.

*Internet Visibility of the Project*

The second Romanian case study was performed by some of the researchers involved in case one, notably B. Sluser and Carmen Teodosiu. The client, however, does not seem to mention the report on any of its pages. An environmental journal mentioned at http://web.ngo.ro/press/persp/pers0898/indexro.htm does no longer seem to exist after 2000.

An report additionally found in case one, entitled "EMS in factories. The introduction of Environmental Management Systems in the metalprocessing and ceramic industry in Romania and The Netherlands," is based on and coauthored by one of the researchers of this project. As noted, it can be retrieved at http://www.fwn.rug.nl/chemshop/cwpub.html.

*Individual Researcher/Mediator Exposure*

The student in case two won the second prize in the Annual Students' Scientific Workshop of the Faculty of Industrial Chemistry in Iasi. The student developed a dissertation thesis that used the data and results from this project. The supervisor of this dissertation, F. Ungureanu, is a well established researcher with the four articles in the *Science Citation Index*, from 1997, 1998, and 2002.

*Conclusion*

The project contains general information about technological processes for yeast fabrication and about wastewaters resulting from this process. It also offers information depicted in literature regarding treatment processes recommend for removal of pollutants from the wastewaters, which resulted from the industrial production of yeast. The impact of wastewaters on the receiving waters was also analyzed, and provided suggestions for improving the environmental situation. As noted, the MA Thesis was awarded with a prize of the faculty.

The NGO used the information presented in the report both for the members of the NGOs as well as information for the local community.

### 3.7.3 Project Vladeni 2000 – Biodiversity Conservation in the Wetland Vladeni (Iasi Country – Romania)

The study was requested by three NGOs: The Romanian Ornithology Society (involved in the monitoring of programs of some wetlands and breeding birds' species current status in Romania), the Romanian Mycological Society and the Society for Ecology (brings together information for people interested in nature knowledge and preservation). The project also involved the InterMEDIU Science Shop and university staff and students from the Faculty of Biology. The research for this project continued previous studies carried out between 1995 and 1998 by the Romanian Ornithological Society, which involved a full biological documentation for a RAMSAR site assessment in the area of the Wetland Vladeni.

The objectives were specified in relation to environmental conservation, ornithological fieldwork objectives, and environmental education objectives. The project took place between January and December of 2000, and all costs were supported by the British Petroleum Environmental Programme, the MATRA project as well by the Faculty of Biology at the University of Iasi.

*Internet Visibility of the Project*

The other university at Iasi is located on the Internet at http://www.uaic.ro/default.php?t=site&lang=EN. "Intermediu" provides four hits in the Romanian pages, but none on English pages. These pages, however, lead to the use of the word "intermediul" and not to information regarding the office.

The NGO, Societatea Ornitologica Romana, is found at http://www.sor.ro/. The project is also retrievable at BP (http://conservation.bp.com/projects/91_proj.asp) and from the NGO at the site http://www.sor.ro/raport.htm. However, the paper given at the Audes Conference in Venice (April 2001) is not visible in the program of the meeting at http://helios.unive.it/~audes6/index.htm.

*Individual Researcher/Mediator Exposure*
C. Gache (SOR-Iasi, Laboratory of Zoology, University of Iasi, Romania) is mentioned as a contributor in the action plan for the corncrake (*Crex crex*) in Europe.  This information is found at http://europa.eu.int/comm/environment/nature/directive/birdactionplan/crexcrex.htm

Searching for the participant Nicoara reveals an established research line, with ten relevant publications. However, a search for Nicoara in the *Science Citation Index* resulted in zero publications.

*Conclusion*
Case three is a thoroughly embedded project on both sides. The local researchers are internationally connected.

### 3.7.4  Conclusions and Recommendations

All three cases are very clear and academic in style. Students are always involved, and the projects are academically embedded. For example, in case one when a questionnaire was constructed, the appropriate academic unit in the social science faculty was consulted by the natural scientists involved in order to warrant the quality control.

The NGOs are often academically oriented.  In case one, the Academic Organisation for Environmental Engineering and Sustainable Development (O.A.I.M.D.D.) acted as the NGO. Case two is based on a question of the Clubul the Ecologie si Turism Moldavia (C.E.T) as a grassroots organization, but the Romanian Ornithological Society is the client in case three.

The Science Shop activities are of high quality and also well received in the environment of the university. Yet, their visibility as Science Shops and as a model of access to the university is still low. The Science Shop activities are subsumed under the institutional organization of the university at the lower end and in terms of scholarly activities they disappear as part of the main structures at the level of the disciplinarily organized departments.

Perhaps, the approach of the university is too traditional. Teodosiu and Caliman (2002) provide an interesting exhibition of Science Shop activities, but the medium of this specialist journal is not sufficient. In our opinion, a responsibility for Science Shop activities should be generated that is different from the interests of the various departments in order to enhance the systematic presentation of results in non-academic terms.

# 4. A Kaleidoscope of University-Society Relations

Danish undergraduate students test the quality of the water in ponds in a suburb because the inhabitants complain about the stink. The students make suggestions to the local municipality which hitherto had neither the technological knowledge nor the resources needed to improve the situation. Secondly, an established scholar in Romania publishes about problems with the quality of surface water caused by a plant at the bend of a river. This project is organized in collaboration with colleagues from the Netherlands, and the work is executed by MA thesis students. Thirdly, a researcher at a public research institute in Valencia researches and advises about the incineration of meat remnants from industrial processes and the potential health risks involved.

These three settings seem to have little in common. However, they were cited as examples of the best practices of Science Shops in reports of the EU-project INTERACTS. What are the communalities among these projects, and how can the results of this wide range of projects be compared? How do these projects relate to community-based research reported in other Science Shops projects or to a Master's thesis about poverty among children in Austria?

Since the emergence of the Science Shops in the 1970s, this model of mediation at the grassroots level has been diffused from the Netherlands within Europe and even beyond (Farkas, 2002; Fischer & Wallentin, 2002; Irwin, 1995; Mulder *et al*., 2001; Sclove, 1995). From its very beginning, the Dutch model was based on coalitions among various groups. Thus one could allow for very different practices reflecting variations among disciplines, political contexts, and institutional organization. The crucial question is how the capacity of public research can be used for solving social and environmental problems, and conversely, how stakeholders in these problems can provide access to questions and domains for the various layers of the university and its respective missions of higher education and scientific research.

For example, some of the Dutch Science Shops (e.g., Utrecht) were heavily engaged in political actions outside the university context, while others (e.g., Amsterdam) considered the shops primarily as an option for institutional reform within the university context. This distinction provides one of the dimensions that can also be observed in the different positions of the Science Shops nowadays: do they wish to operate given a university context or do they additionally try to change the academic arrangements? Answers to these questions can be expected to vary between more established disciplines (e.g., chemistry) and more action-oriented fields such as women's studies.

The Science Shop in Innsbruck (the FBI Institute), for example, profiles itself as a knowledge base and an independent center of expertise on the side of action groups and NGOs, while Science Shops elsewhere in Austria mainly provides university students with possible topics for their Master's theses. Note that what counts as relevant outputs in these very different configurations may be different because the institutional roles and the professional expectations are structurally different.

## 4.1 What is mediated by the mediators?

The INTERACTS project consortium (Fischer & Wallentin, 2002) argued in the state-of-the-art report about Science Shops in Europe (WP3) that the mediating role itself is under pressure to change when the relations between the two sides of the mediation are changing over time. Four waves of Science Shop work were distinguished:

1.  The initial wave of the foundation of mainly Dutch Science Shops during the 1970s. These developments were initiated from within the university system by coalitions of progressive staff members and student activist movements. The terms of the debate were given by the science policy discourse about "democratization," that is, access to the scientific knowledge bases of society.
2.  A second wave (in the 1980s) was strongly interwoven with the further institutionalization of alternative movements like the "Bürgerinitiative" in Germany. These groups in civil society (i.e., outside academia) needed to develop their knowledge base and sometimes turned for assistance to the university. Furthermore, some of these NGOs recruited membership among students and university staff members. Thus, a common perspective could sometimes be developed.
3.  A third wave has been based on the increased awareness of the need to build social capital and to fight exclusion mechanisms in the post-Cold War societies of the 1990s. From this perspective, the network function across institutional boundaries becomes an objective in itself. Science Shops provide a model for engaging the university in non-economic objectives with groups which were hitherto excluded from these knowledge-intensive domains.
    While the first wave was mainly grounded in the movements of students and critical staff at the universities and the second was mainly grounded in civil society, during the third wave one has tried to bridge this gap on the basis of professional considerations. This wave is therefore mainly driven by social scientists. The community-based research at Liverpool (paragraph 3.6 above) can be considered as an example of this latter type of project (Hall & Hall, 1996). Through the disciplinary dimension of community-based research, relations with similar groups in the U.S.A. and Canada functioned as an important resource for the movement during the 1990s (e.g., Sclove, 1995).
4.  A fourth wave can be distinguished in the countries of Eastern and Middle Europe and perhaps we can also place the Science Shops in South Africa and Third World countries in this context (Mulder *et al.*, 2001). Groups and organizations in these countries could use the model of Science Shops during the reconstruction of their economies and their societies. In such situations one may locally be able to explore the synergetic potentials of new arrangements (Bunders & Leydesdorff, 1987). In Spain—one of the case studies of the INTERACTS project—one would expect to find an early example of this construction of civil society after the Franco period. The knowledge base of society can be used as a cultural resource for the reconstruction of institutional arrangements.

Bridging mechanisms at the institutional level like Science Shops can be useful in developing the social integration. Functionally, these intermediaries may also provide a counterbalance and legitimation in a context where more commercially oriented technology transfer and science parks are supported for economic reasons. The interfaces of the academic community with the economy and with processes of political decision-making in the public sphere cannot be expected to coincide in a complex and plurifom society. On the contrary, one expects tension between the public and the private domains over matters of access and principles of appropriation. Therefore, one would expect the mechanisms of transfer also to be different.

During the last two waves of the 1990s, professionalization of the transfer of knowledge has become crucial because the creation of social capital is based on the generation of trust. In this context, the quality control of projects can no longer be left to undergraduates and student volunteers at lower levels of the organization. The institution itself has to make a visible commitment if one wishes to contribute to the objective of generating social capital. Quality control can be made the responsibility of specific departments and disciplines (for example, of sociology in the collaboration between the two universities at Liverpool) or the Science Shop itself can be further professionalized (like the extra-academic Science Shops in Bonn and Innsbruck).

Thus, the Science Shops have evolved in terms of the forms that are sustainable in their present environments, and furthermore in terms of the specific forms of integration that have been functional during the last decades. It is therefore not surprising that there is a flux of new entrances and exits among Science Shops. A reflection of the current state may thus provide us with a variety of elements that are crucial for "a new social contract between science and society" from a perspective not exclusively economic (Caracostas & Muldur, 1998; European Commission, 2001; Gibbons *et al.*, 1994). Science Shop practices offer a point of crystallization where the latent demand for public access to science and technology can become visible, not only as an intellectual interest (as in a science museum), but as an interest rooted in social structure. The clients of the Science Shops formulate a need to obtain access. These needs are socially structured and therefore the sociologist expects a kaleidoscope of recognizable formats rather than an amorphous grey mass (Leydesdorff, 1993).

## 4.2   A new social contract of science

In principle, the Science Shop case studies may thus teach us about the elements that can be made functional when drafting a new social contract between science and society. This process is not one-sided as a supply-driven client relation because the knowledge base is expected also to structure the configuration. Knowledge-based insights and innovations are not like commodities freely available on an anonymous market, but they can be considered as highly specific (because codified) channels of communication between science and the public. For example, a client may turn to a university or Science Shop with a specific question for which no answer is yet available. The question may, however, be very relevant from the analytically different perspective of a specific research program.

How this mediation can be institutionalized is another issue; the institutional framing requires a further reflection on the local contexts. At the institutional level one can thus expect specificity in the windows of communication to be further developed. For example, within the higher education system teachers have an institutional need to provide students with relevant and interesting research questions for their Master's theses. This point of entrance for client topics can easily be recognized as a place where the higher-education system opens up for a specific form of delivery of expertise to social groups. Unlike Ph.D. projects, theses at the Master's level are usually not expensive (because of, for example, internship relations). Furthermore, if sufficiently supervised the results can be used as sources of legitimation in a political discourse, while in technology transfer economic considerations require a higher degree of reliability and planning. MA students, for example, often do not finish within the envisaged time planning.

But this is not the only perspective and window. For example, the third Spanish case shows the need of trade-unions to build up in-house expertise that is backed up by the research system, but not necessarily at the university level. Their extra-university expertise calls here upon a public research institute to exercise a quality-control function for the relevant knowledge. In a knowledge-based society, the organization of scientific knowledge can be expected to be needed in domains other than academia. In such cases, whether organizations and groups in the civil society will able to turn to academia with their questions depends on the development of relations of mutual trust. Although the Science Shops provide a low threshold, NGOs that have developed an internal knowledge base may prefer to discuss and negotiate with scientific departments and institutions without intermediation. This is particularly likely if the NGO involved has in-house academics who already belong to their own intellectual networks.

## 4.3   The macro-context

Let us first summarize from a historical perspective how the four waves of the institutionalization of Science Shops can be considered as results of changes in the relations between science and society. From the perspective of the local institutions these changes in the organization of society took place at a global level. However, changes in these regime inflict on the trajectories of the institutions carrying the interfaces because they affect the conditions of the institutionalization (Giddens, 1984).

The first wave and the student movement can be recognized with hindsight as a product of the welfare state that emerged in Europe in the 1960s. The student movement was culturally motivated to use science for purposes other than economic advancement. This was often expressed with the slogan of the 1968 revolt "*l'imagination au pouvoir*" ("power to the imagination"). The experiments of Science Shops in the 1970s can be compared with other attempts to generate bridges between a left-wing (socialist) appreciation of science and technology, and further reflection among intellectuals about cultural changes in the role of scientific knowledge in society. These roles were changing because knowledge production was increasingly organized and controlled (Habermas, 1968; Whitley, 1984). National science and technology policies were in this period still very much under construction.

For example, in the 1970s the Labour Minister of Industry in the U.K., Tony Benn, endorsed the so-called Industrial Workers Plans (Cooley, 1980). The German Ministry of Science and Technology (BMFT) launched an ambitious research program for the "Humanization of Labour," and alternative product designs involving users in an early stage became almost a benchmark of Scandinavian quality (L.O., 1982; Leydesdorff & Van den Besselaar, 1987a). The French "colloque national" on science and technology in 1981 can similarly be considered as part of this development (Leydesdorff and Van den Besselaar, 1987b; Vavakova, 2000). Richta *et al*. (1968) report about similar discussions in Czechoslovakia concerning the role of scientifically organized knowledge in the economies of Eastern Europe during the "Spring of 1968."

Perhaps, one can ideal-typically identify this model as "old left" (marxist or neo-marxist) versus the "new left" or "green" models of the 1980s. The first wave focused on institutional reform, while the second model also had an anti-institutional flavour. The environmental movement considered science and technology as themselves part of the problems caused by industrialization, and therefore science could not so easily be expected to provide solutions to problems formulated by the environmental movements of that time. In practice, however, the two models have continuously been recombined, but for the sake of analytical clarity it can be useful to highlight these distinctions.

The second model emerged in relation to the decline of the industrial model for economic development following the oil crises of 1973 and 1979. In the advanced economies, the emphasis shifted in this period from science policy to technology and innovation policies (OECD, 1980). While low-wages countries were able to outcompete the OECD countries in terms of labour and raw materials, the advanced industrial systems increasingly turned to knowledge-based innovations to maintain their competitive edge. During the early 1980s, this turn led to a rethinking of industrial policies under a neo-liberal regime that was strategically oriented towards "re-industrialization" (Rothwell & Zegveld, 1981). The environmental movement was thus pushed into the position of having to defend the environmental legislation of the 1970s.

Although the alternative movements remained ambivalent about scientific rationalization, some elements of the environmental movement found shelter in the university as a breeding place for cultural reflection and for the development of alternative models (Beck, 1992). University departments of environmental sciences had been institutionalized successfully during the 1970s, but at many places these "interdisciplinary" units remained vulnerable to budget cuts given the new regime that focused, for example, on the development of entrepreneurial activities like "biotechnology" and ICT.

Our case materials provide more examples from this second wave than from the older model, but sometimes one can also analyze the cases in terms of combinations of all three movements: the alternative movement of the 1980s with its strong focus on environmental issues, the cultural reform demanded by the student movement (e.g., still very much alive in Berlin), and the trust in science and technology as potential forces of transformation and enlightenment viewed from a more traditional perspective. As noted, a fourth element has to be added because of the more recent focus on the use of bridging institutions for the generation of social capital. In this dimension, a disciplinary perspective within the social sciences, notably, a focus on "grounded theorizing" (Glaser, 1992; Glaser & Strauss, 1967), found an institutional form in activities that are very akin to the Science Shop model (Hall & Hall, 1996).

With hindsight, one is now able to appreciate these different dynamics as cultural (sub)dynamics that can continuously be recombined in the complex evolution of social relations (Luhmann, 1984). The case studies illustrate how these different resources and repertoires are parts and parcels of European culture. For example, the "old" model of anti-capitalist opposition is still important in one of the Spanish cases. In the Austrian case studies the concern for groups marginalized by capitalist development can also be considered as a driver, but now with a focus on "inclusion." All the cases can be considered with hindsight as examples of the construction of social capital.

The emphasis on the value of networking itself has been typical for the 1990s. Fukuyama (1999), for example, discussed the crisis of modern society in terms of the erosion of social capital (Putnam, 2000). From this perspective, network formation can be considered as an objective in itself because networks provide a mechanism for preventing social exclusion. The social networking, however, cannot be achieved by deliberate policies at the national level, but has to be made possible within civil society, for example, in terms of new interactions among existing institutions.

## 4. 4   Shaping public systems of innovation

The gradual decrease of identity at the national level was reinforced in Western Europe by the end of the Cold War, the subsequent disappearance of the Soviet Union and the reunification of Germany, and—last but not least—the emergence of the Internet. The concept of the "new economy" became virtually synonymous with a knowledge-based economy that is innovation-driven and globalized. While globalization can be perceived as a threat, innovation can be turned into a celebration of community formation because the new products and processes have to be accepted and disseminated locally. New conditions enable and allow people to cross bridges across ethnic groupings and over institutional (and national) divides (Leydesdorff & Etzkowitz, 1998). This spirit of new options because of changing competitive conditions across borders can be considered central to the EU project itself.

From the perspective of prioritizing the generation of social capital, the role of the mediation changes again. The networks generate knowledge endogenously in terms of consensus and dissensus formation, exchanges of convictions, rational expectations, and arguments. High-quality communication across borders becomes more important than providing expertise and counter-expertise in an oppositional mode. The university can be expected to play a constructive role in these exchanges, while the partners can accept that the university has to cultivate roles like guarding the quality of scientific communication at the global level and providing society with qualified personnel through higher education. Given these global objectives of the university, the institutional parameters of the operation can be debated and reconstructed among the partners in terms of locally optimal conditions (Van den Daele *et al.*, 1977).

For example, some universities have been developed with venture capital during the 1990s into "entrepreneurial universities" (e.g., Chalmers University in Gothenburg; cf. Clark, 1998), while others—for example, those located in less-favoured conditions—have attempted to act as regional innovation organizers (Etzkowitz *et al.*, 2000; Leydesdorff *et al.*, 2002; Pires & Castro, 1997). The intellectual tasks and missions of the university require the appreciation of additional sources of funding, but sometimes also counter-balancing measures against strong pressures to commercialize. The saliency of the university within these processes is a consequence of its need to be reflexive on the labour market and therefore adaptive to the conditions in which it has to organize local niches of knowledge-intensive development. The university provides these environments in turn with new (and sometimes counterintuitive) insights. Unlike the systems in its environment, however, the main mover of this system is the continuous flux of students, i.e., human resources. The latter have to be provided with opportunities for further academic careers, functions in the civil service, and perspectives on industrial and entrepreneurial activity. In other words, the university's mission is to recombine these functions (Brewer, Gates, & Goldman, 2001).

## 4. 5   Normative and cognitive orientations in social capital

The social systems of science and the clients of the Science Shops meet at interfaces. However, the two groups can be expected to use very different repertoires and to provide different meanings to the relevant interfaces. First, the two sides are differently organized, potentially (but not necessarily) in terms of institutions. Second, one expects the horizons in the communication to be developed very differently among the partners. Therefore, the Science Shop functions both as an institution *and* as a mechanism for translation. The relative emphasis on the institution and on its functions can be expected to vary among Science Shops. The institutions, however, condition how these functions can be fulfilled.

From the perspective of mediation, the bridging function of the Science Shop as an institution can be considered as providing an infrastructure for the further development of the knowledge base within and among the three partners involved. These three partners are the clients on the demand side, the researchers and students on the supply side, and the mediators at the level of the Science Shop. One cannot expect a one-to-one correspondence between the types of knowledge developed and these institutional roles, but instead one would expect a further differentiation. The parties bring different elements relevant to the knowledge production process at the interface, and they expect different elements in return.

For example, the clients may be able to offer access to domains which are not easily open to questioning without their help (Zaal & Leydesdorff, 1987). The staff and students at the university are experts on the side of the methodologies and techniques to be used. The theoretical dimension sets other constraints which can be expected to vary among disciplines and specialties. For example, using the "grounded theory" approach of social scientists, one would give priority to the reconstruction of the perspective of the subjects under study. The aim is to improve the communicative competences of the partners qualitatively by providing them with (potentially counterintuitive and therefore emancipatory) insights. The same report, however, also evaluates and potentially improves the intervention. The reconstruction may provide the different participants with options for change and thus for solving puzzles. The reflexive formulation of these perspectives can help to enlighten other levels of policy making, but this will be considered as an indirect effect by scholars working from within this theoretical tradition.

Issues of health and safety are often of a far more objective nature. When an action group comes to a Science Shop with a worry about their environment (for example, the contamination of water), this group does not expect first to be counselled by the Science Shop mediator about why these people are worried subjectively. The NGOs in this case may wish to use the results at other levels of policy-making. The main interest may be in participation in decision making, and not in the research process itself. Thus, the nature of the participation of the citizenship can be expected to vary with the cognitive nature and the social functions of the research questions involved.

The resulting reports and other outcomes have different functions for the three partners involved and these contents have to be "translated" in terms of their relevance for the wider environments. These communication processes involve asymmetries in the understanding that provide sources of potential conflict. The normative integration on each side has to be balanced by rationalized differentiations so that these immanent conflicts can be handled. If the normative integration were to fail on either side, the exchange would become risky and the intermediation could be expected to fail because the partners would tend to withdraw using their own frameworks. If the rationalized differentiation fails, however, distrust can be expected to emerge because cognitive expectations (as different from normative ones) are damaged. The results of scientific investigations easily lead to counter-intuitive conclusions that need extensively to be explained.

The two mechanisms (normative and cognitive expectations) operate at different levels and can disturb a project for very different reasons and with an interaction effect. The failure of normative expectations in the communication functions differently from the failure of cognitive expectations. The latter can be expected to generate a breakdown of trust in the relation, while the former generates anxiety and therefore failure to communicate. A Science

Shop has to operate and to succeed at these two levels at the same time. Analytically, this leads to a drive towards professionalization.

The most extreme case of professionalization and institutional independence is illustrated by the report about the Science Shop in Bonn. This Science Shop not only functions as an independent association, but has also internalized substantive expertise about the mediation and factors relevant to it (like funding). The Shop publishes independently, and operates outside academia as a consultancy agency. Most of the Science Shops, however, are strongly related to universities and do not publish independently because publication is precisely the scholarly task of the scientists (and students) involved. Externally, the main function of the Science Shop is then to alert the press and to stimulate the publication of brochures. Brochures can be used as pragmatic versions of the (scientific) reports by the client groups.

## 4. 6   Communication of the Mediation

Science Shops are sometimes insufficiently aware that the reports have only the status of grey literature within academia. They are not considered as scientific output that can be submitted to scientific journals. Within the scientific production process of scholarly publications, the reports therefore tend to disappear. Further reflection is necessary in scientific publications and higher education. In the case of Science Shop projects, this reflection is often further developed within the scientific institution, but the reflexive communication is then no longer attributed to the Science Shop (Zaal & Leydesdorff, 1987). It contains another selection mechanism generated by using the codes of the research system and/or the higher-education system.

Given their increasingly important role in social capital formation, the Science Shop reports and corresponding MA theses deserve to be archived. Archiving and publication at the Internet should further be developed in current Science Shop practices. In some cases, the reports are suspiciously absent, that is, to such an extent that one wonders what happened with these reports. In other cases, the materials are linked to assistants and professors who may have disappeared from the institutional presentation of materials on the Internet because they left the institution (for one reason or another). Because a number of the researchers who acted as supervisors were not on a tenure track, this change of position and therefore visibility could have been anticipated by the local Science Shops.

In the past one has tried to organize the documentation of the Science Shops centrally. For example, the SciPas project contained an ambitious vision of the creation of a central database by the Loka institute. This database was fully programmed and brought on-line (at http://www.livingknowledge.org), but it was never filled with reports by the different Science Shops. With hindsight, this project may have been too ambitious and too centralized. The Science Shops fulfill functions for the respective universities in their local contexts and it is therefore often important that the reports are profiled as output of the respective universities. The local pages can be mirrored (translated, and perhaps classified) at the supra-national level of a centralized database, but only the mirroring and the further elaboration of search facilities on that basis is then a task for this project. As long as the lower-level is not well in place, however, the higher-level aggregation can be expected to remain incomplete.

The Science Shops should be encouraged to make reports available (as some of them are) as files on the Internet. This could be standard practice and one of the criteria for further funding. It provides researchers, students, and clients with points of reference in their practices, and publication on the Internet can be expected to provide more access and recognition from various sides (Lawrence, 2001). The availability of reports and the active updating provides opportunities to claim the credit for an innovation at a later stage, even if the effects of the new insights are somewhat disappointing in the short term. For example, if in a later stage (e.g., after the next elections) the municipality of Frederikssund should decide to clean the stinking pools in their village, the reconstruction would become partly attributable as credit to the students who took the initiative. It would be impossible to ignore this link if the reports were properly archived.

What the reports mean on either side of the interface can again be expected to be different. We found several cases where students who wrote a Master's thesis in the framework of the Science Shop were among the best and were provided with career opportunities at their respective universities. We found also cases where the students left academia, but were included as coauthors in a later publication by the supervisor. The winning of prizes and awards by both students and scholars involved in Science Shop work is remarkable. In the Berlin cases this mechanism of recognition and appreciation seems well developed.

We would advise the Science Shops to learn from these mechanisms. One can provide the authors of reports with serious certificates (e.g., from the university) on a yearly basis. A number of "best practices" can also be defined as: "best student paper," "best report," "best advice," etc. The jury could be staffed by a board-like committee where clients, administrators, and scholars (university staff and/or externally) meet to discuss the results of the past year with the purpose both to provide recognition for the students and scholars involved and to make recommendations for improvements in the quality of the mediation.

## 4. 7   National and disciplinary differences

Two major structural dimensions for comparison among the cases are provided by national differences among Science Shop practices and the disciplinary affiliations of the researchers. Perhaps, with the exception of Spain where the Science Shop is not yet itself a concept used for the mediation, the common origin of the discourse about Science Shops in the various European countries is recognizable. These activities seem to attract highly motivated, culturally advanced, and socially engaged students and young scholars who are seeking to make careers that are intellectually and socially meaningful. The Science Shops provide and generate social capital in terms of relevant networks first of all for the researchers involved. These projects can perhaps be considered as a distributed format of the new social contract between the universities involved and their environments.

The mediating role is widely recognized, to such an extent that often the Science Shops can turn routinely to institutionalized offices within the departments that help students for internships, etc. These provisions are standardized in departments in the social sciences more than in the natural sciences. Some of the Shops are heavily related to specific departments in applied natural science areas. Thus, the disciplinary categorization can become more important than the national one for the mechanisms that prevail in the mediation.

For example, while differently structured in terms of their institutional organization, the Science Shops of Vienna and Liverpool provide in some respects similar services to both MA students and clients. While the team in Liverpool is highly focused on its disciplinary orientation as social scientists, the wider range of supervisors in the Vienna cases are intellectually organized in similar frameworks. These authors may not know each other because most of the publications by the Austrian scholars are in German, but the orientation in terms of "grounded theory," appreciation of non-economic objectives, supervision of students, etc., are comparable.

A second group of shops is focused on "environmental issues," but here a further distinction can be made. The Science Shop activities in Romania, Denmark, and probably a number of the Dutch shops (which were not included among the cases studied in this project) can be subsumed under the heading of "environmental engineering and management" (Teodosiu & Caliman, 2002). The structural position of these problems in Western and Eastern Europe is different, but this field is still an important driver of Science Shop activities in both research and higher education.

Although environmental issues have been a driver in organizing Science Shops in Germany since the 1980s, the German case studies have in common with the Spanish that they are less oriented towards higher education, and more towards professionalization at the level of the scientific staff. In these two nations, Science Shop questions offer access to intellectual domains that are experienced as challenging by individual scholars. The professional activities in these two systems seem institutionalized. The Innsbruck institution FBI carefully wishes to prevent its further institutionalization, but it can still be compared as a professional unit on these dimensions.

In summary, three substantive foci can be distinguished among the twenty-one case studies:

1. "environmental engineering and management" both as a disciplinary profession and in terms of higher education (Teodosiu & Caliman, 2002);
2. social work, its problems, and its evaluation aimed at the improvement of intervention strategies (Hall & Hall, 1996);
3. professional expertise in the case of problems that cannot easily be disciplined (Schweighofer-Brauer *et al*., 2002).

These intellectual foci interact with two types of structural relations that can be maintained from the side of the university: a relation based on the higher-education function of this institution, and one based mainly on its research function. Questions of clients of Science Shops provide the higher-education system with topics, for example, for writing Master's theses. The university can consider using these topics and theses as starting points for the institutionalization of new research lines, but this is currently not a standard practice. In most cases, the individual students are successful in using the advantages of their specific projects for their further career either within or outside the home university.

The topics are primarily brought into the system at the level of research if the interface between the clients and the Science Shops has been further professionalized. This professionalization can occur on either side of the interface. In such configurations, the topics are already provided with a specific interpretation in terms of a scientific specialty before the research process has begun. If the professionalization is developed on the side of the client, then specific networks are used for the mobilization of expertise. But if the professionalization has mainly taken form on the side of the Science Shop, the latter tends to function as a consultancy operation with a low threshold for economically weak but politically urgent demands.

In the Romanian cases we found a more integrated approach in terms of the higher-education and research functions of the university. This may relate to the specific phase of the development of civil society and the discipline of environmental engineering in Romania. In this country, networks seem to be mobilizable across the institutional divides of established institutions. The system can perhaps be considered as less differentiated and fragmented than in some of the Western European countries at this stage. Perhaps this emerging form of organization can be developed into a more stable advantage of the transitional economies.

If the university would like to profit from societal input both at the level of higher education and at the level of research, communalities in the interfaces of research and higher education with the university environment can be further developed. We have indicated a few of such common points, such as the establishment of rewards for best practices. Such structures may have to distinguish between social relevance and scientific quality. A university may wish to establish a standing committee at the level of the board that investigates the potentials for further development at the interface with the surrounding society more systematically and in terms of both research and higher education. The European Commission could provide universities that wish to move in this direction with startup subsidies (on a competitive basis) in order to strengthen the implementation of Action 21 of its Science & Society Action Plan.

## 4. 8  Conclusion

The European Commission expressed itself as follows in the *Science & Society Action Plan* of December 4, 2001:[27]

> "There are in Europe various types of Science Shops close to the citizen in which science is placed at the service of local communities and non-profitmaking associations. Hosted by universities or independent, their common feature is that they answer questions from the public, citizens' associations or NGOs on a wide variety of scientific issues. The first Science Shops were opened in the Netherlands in the 1970s and the idea was then taken up by about 10 other countries throughout the world. There are now over 60 Science Shops in Europe, mainly in the Netherlands, Germany, Austria, the United Kingdom and France.
>
> The diversity and scope of questions is such that the most successful centres are having difficulty in satisfying demand. The Science Shops would gain from getting together, with the aid of the Commission, to pool their resources, their work and their experience."

A series of three European projects has now been funded by the Commission. INTERACTS (2001-2003) followed upon an extensive inventory by SciPas (1991-2001), and this year (2003) the above announced network ISSNet was also created. SciPas focused on making an inventory of ongoing activities. The main purpose of INTERACTS, however, was to analyze the practices of Science Shops substantively by using in-depth case studies. The results of this project enabled us to understand how the transfer of science and technology works at the grassroots level.

The case studies show that despite the variation in terms of nations, disciplines, institutional settings, etc., the Science Shops have developed a common language of mediation between citizen groups and the public sphere. This common language has evolved in local niches as best practices of mediation. The communication of Science Shop mediation adds another layer to the practical mediation itself. The comparison among the case studies allowed us to distinguish the institutional integration between the higher-education function and the research function of the university. This distinction is perhaps more important than the focus on national differences.

The tasks of the Science Shops in recombining normative concerns with analytical perspectives could further be explored and the inherent tensions in this type of work made visible. It seems obvious that wherever these mechanisms are successful in solving the puzzles involved, they can be expected to remain fragile. The Science Shops operate at interfaces which are not continuously needed from the perspective of the institutions. However, these interfaces may be crucial for the development of a knowledge-based society from a system's perspective. The translation of clients' concerns and demands into the system and the feedback of supply by research and higher-education legitimates the latter and this mediation deeply involves public audiences because their substantive demands are taken seriously. Academic freedom can thus be appreciated more fully as a societal resource.

---

[27] The full text of the *Science and Society Action Plan* of the European Commission can be retrieved at ftp://ftp.cordis.lu/pub/rtd2002/docs/ss_ap_en.pdf